\documentclass{aa}  

\usepackage{graphicx}
\usepackage{txfonts}
\usepackage{natbib}
\bibpunct{(}{)}{;}{a}{}{,}

\usepackage{txfonts}
\begin{document}

   \title{The Gaia-ESO Survey: Kinematic structure in the Gamma~Velorum
     cluster\thanks{Based on observations collected with the FLAMES spectrograph at VLT/UT2 telescope (Paranal Observatory, ESO, Chile), for the Gaia- ESO Large Public Survey (188.B-3002).}}
%

   \author{R.~D.~Jeffries
          \inst{1}
          \and
          R.~J.~Jackson \inst{1}
          \and
          M.~Cottaar \inst{2}
          \and
          S.~E.~Koposov \inst{3}
          \and
          A.~C.~Lanzafame \inst{4}
          \and
          M.~R.~Meyer \inst{2}
          \and
          L.~Prisinzano \inst{5}
          \and
          S.~Randich \inst{6}
          \and
          G.~G.~Sacco \inst{6}
          \and
          E.~Brugaletta \inst{4}
          \and
          M.~Caramazza \inst{5}
          \and
          F.~Damiani \inst{5}
          \and
          E.~Franciosini \inst{6}
          \and
          A.~Frasca \inst{7}
          \and
          G.~Gilmore \inst{3}
          \and
          S.~Feltzing \inst{9}
          \and
          G.~Micela \inst{5}
          \and
          E.~Alfaro \inst{8}
          \and
          T.~Bensby \inst{9}
          \and
          E.~Pancino \inst{10}
          \and
          A.~Recio-Blanco \inst{11}
          \and
          P.~de~Laverny \inst{11}
          \and
          J.~Lewis \inst{3}
          \and
          L.~Magrini \inst{6}
          \and
          L.~Morbidelli \inst{6}
          \and
          M.~T. Costado \inst{8}
          \and
          P.~Jofr\'e \inst{3}
          \and
          A.~Klutsch \inst{7}
          \and
          K.~Lind \inst{3}
          \and
          E.~Maiorca \inst{6}
          }

   \institute{Astrophysics Group, Keele University, Keele, 
      Staffordshire ST5 5BG, UK\\
              \email{r.d.jeffries@keele.ac.uk}
         \and
             Institute for Astronomy, ETH Zurich,
             Wolfgang-Pauli-Strasse 27, 8093, Zurich, Switzerland
          \and
Institute of Astronomy, University of Cambridge, Madingley Road,
Cambridge CB3 0HA, United Kingdom
\and
Dipartimento di Fisica e Astronomia, Sezione Astrofisica, Universit\'a di Catania, via S. Sofia 78, 95123, Catania, Italy
\and
INAF - Osservatorio Astronomico di Palermo, Piazza del Parlamento,
Italy 1, 90134, Palermo, Italy 
\and
INAF - Osservatorio Astrofisico di Arcetri, Largo E. Fermi 5, 50125,
Florence, Italy
\and
INAF - Osservatorio Astrofisico di Catania, via S. Sofia 78, 95123 Catania, Italy
\and
Instituto de Astrof\'{i}sica de Andaluc\'{i}a-CSIC, Apdo. 3004, 18080,
Granada, Spain
\and
Lund Observatory, Department of Astronomy and Theoretical Physics, Box
43, SE-221\,00 Lund, Sweden
\and
INAF - Osservatorio Astronomico di Bologna, via Ranzani 1, 40127,
Bologna, Italy
\and
Laboratoire Lagrange (UMR7293), Universit\'e de Nice Sophia Antipolis,
CNRS,Observatoire de la C\^ote d'Azur, BP 4229,F-06304 Nice cedex 4,
France
}

   \date{accepted January 20th, 2014}

 
  \abstract
   {A key science goal of the Gaia-ESO survey (GES) at the VLT
     is to use the kinematics of low-mass stars in young clusters and star forming
     regions to probe their dynamical histories and how
     they populate the field as they become unbound. The clustering of low-mass stars around the
massive Wolf-Rayet binary system $\gamma^2$~Velorum was one of the first GES targets.}
   {To empirically determine the radial
velocity precision of GES data, construct a kinematically unbiased
sample of cluster members and characterise their dynamical state.}
   {Targets were selected from colour-magnitude diagrams and
     intermediate resolution spectroscopy was used to derive radial velocities and
     assess membership from the strength of the Li~{\sc
       i}~6708\AA\ line. The radial velocity distribution 
     was analysed using a maximum likelihood technique that accounts
     for unresolved binaries.}
   {The GES radial velocity precision is about 0.25\,km\,s$^{-1}$ and sufficient to resolve
     velocity structure in the low-mass population around
     $\gamma^2$~Vel. The structure is
     well fitted by two kinematic components with roughly equal numbers
     of stars; the first has an intrinsic
     dispersion of $0.34\pm 0.16$\,km\,s$^{-1}$, consistent with
     virial equilibrium. The second has a broader dispersion of
     $1.60\pm 0.37$\,km\,s$^{-1}$ and is offset from the first by $\simeq
     2$\,km\,s$^{-1}$. The first population is older by 1--2\,Myr based
     on a greater level of Li depletion seen among its M-type stars and
     is probably more centrally concentrated around $\gamma^2$~Vel.}
   {We consider several formation scenarios, concluding that the
    two kinematic components are a bound remnant of the original,
    denser cluster that formed
    $\gamma^2$~Vel, and a dispersed population from the wider Vela OB2
    association, of which $\gamma^2$~Vel is the most massive
    member. The apparent youth of $\gamma^2$~Vel compared to the older
    ($\geq 10$\,Myr) low-mass
    population surrounding it suggests a scenario in which
    the massive binary formed in a clustered environment 
    after the formation of the bulk of the low-mass stars.}

   \keywords{stars: pre-main-sequence -- stars: formation -- stars:
     kinematics and dynamics -- open clusters
and associations: individual: $\gamma^2$ Velorum}

   \maketitle
%

\section{Introduction}

The Gaia-ESO survey (GES) is employing the FLAMES
multi-object spectrograph (Pasquini et al. 2002) 
on the VLT UT-2 (Kueyen) telescope to obtain
high quality, uniformly calibrated spectroscopy of $>10^{5}$ stars in
the Milky Way (Gilmore et al. 2012; Randich \& Gilmore 2013).  
The survey covers stars in the
halo, bulge, thick and thin discs, as well as in star forming regions
and clusters of all ages. Samples are chosen from photometric surveys
with the overarching aim of characterizing the chemical and
kinematic evolution of these populations.  The survey will provide a
rich dataset which, when combined with proper motions and parallaxes
from the forthcoming Gaia mission (Perryman et al. 2001), will
simultaneously yield 3D spatial distributions, 3D kinematics, chemical
abundances and astrophysical parameters for large numbers of
representative stars.
 
One of the key science drivers of the survey is probing the formation
and subsequent dissolution of young clusters and associations using the
kinematics of their constituent low-mass stars.  It is often claimed
that most stars form in clusters, but a comparison of the observed
number of clusters embedded in their natal gas with older, gas-free
open clusters suggests that 90 per cent of clusters must either start
in an unbound state or become unbound during this transition (Carpenter
2000; Lada \& Lada 2003). The heating and subsequent
expulsion of gas by ionising radiation, winds or supernovae in clusters
containing high-mass stars, but with relatively low star forming efficiency, is
likely to unbind a significant fraction of their stars and possibly
disrupt the whole cluster (Tutukov 1978; Hills 1980;
Goodwin \& Bastian 2006; Baumgardt \& Kroupa 2007; Bastian 2011).  OB
associations have a similar stellar content to young clusters but are
of much lower density and may be the unbound remnants or halos of
clusters after the gas expulsion phase (Kroupa et al. 2001; Clark et
al. 2005). Alternatively, it is possible that the importance of
clustered star formation has been overestimated and that stars are
formed hierarchically in environments with a wide range of initial
densities; bound clusters are formed from the densest regions whilst
associations formed at low densities and were possibly 
never bound to begin with (e.g. Bressert et al. 2010; Bonnell et
al. 2011; Kruijssen et al. 2012).

The key to understanding the past and future evolution of clusters and
associations lies in careful measurements of the positions and
velocities of their constituent stars. Gaia will ultimately yield very
precise tangential motions, but observations of radial velocities (RVs)
and RV distributions in cluster and association populations can be used
to assess membership, probe the current dynamical state, search for and
parametrise binary populations, and investigate spatially coherent
velocity gradients or substructure that might give clues to the initial
conditions or reveal multiple populations (e.g. Jeffries et al. 2006;
F{\H u}r{\'e}sz et al. 2006, 2008; Brice\~no et al. 2007; Sacco et
al. 2008; Maxted et al. 2008; Tobin et al. 2009; Cottaar et al. 2012a).

The GES began on 31 December 2011 and will be completed over the
  course of $\simeq 5$ years. Many important results regarding the
  issues discussed above will emerge from a homogeneous analysis of the
  $\simeq 30$ young ($< 1$\,Gyr) clusters that will eventually be targeted, but
  significant progress can be made before then
  because the data for individual clusters are 
  usually collected in one observing season, and these datasets
  can serve to refine and test analysis
  techniques. The first "cluster"
target was the collection of young, low-mass stars around the massive
WC8/O8III binary system, $\gamma^2$ Velorum (HD 68273, WR11; Smith 1968; Schaerer, Schmutz \&
Grenon 1997). This binary, with a 78.5-day period and eccentricity of 0.33
(North et al. 2007),  is the most massive member of the common
proper motion Vela OB2 association, consisting of 93 early-type
candiate members spread over 100 square degrees (de Zeeuw et al. 1999). The
Wolf-Rayet and O-star components have current
masses of $9\,M_{\odot} + 30\,M_{\odot}$ (de Marco \& Schmutz 1999), but
initial masses of about 
$35\,M_{\odot} + 31.5\,M_{\odot}$ (Eldridge 2009). A further common
proper motion component of the system, $\gamma^1$~Vel
(HD68243), is separated by 41 arcsecs to the south-west and is itself a
multiple system containing a close SB1 binary with a B2III primary
(Hern\'andez \& Sahade 1980) and a tertiary at 0.037 arcsecs that is 1.8
mag fainter (Tokovinin, Mason \& Hartkopf 2010). A
surrounding association of low-mass pre-main-sequence (PMS) stars was
first identified by
Pozzo et al. (2000) by virtue of their strong X-ray emission. A consideration of the
colour-magnitude diagram of the X-ray sources and their concentration
around $\gamma^2$ Vel, led Pozzo et al. to conclude that the PMS stars
were coeval with the massive binary at an age of $\sim
4$\,Myr, and at a distance of 350--400\,pc. This distance is
approximately consistent 
with the revised Hipparcos parallax-based distance to $\gamma^2$~Vel of
$334^{+40}_{-32}$\,pc (van Leeuwen 2007) and with interferometric
determinations of $368^{+38}_{-13}$\,pc and $336^{+8}_{-7}$\,pc by
Millour et al. (2007) and North et al. (2007) respectively.

Jeffries et al. (2009, hereafter J09) made a further study of the relationship between
$\gamma^2$~Vel, Vela OB2 and the X-ray-active PMS stars using a 0.9
square degree $BVI$ photometric survey centred on $\gamma^2$~Vel,
supported by XMM-Newton X-ray observations and some fibre spectroscopy
of solar-type candidate association members. They confirmed that the
PMS stars are strong X-ray emitters and are spatially concentrated
around $\gamma^2$~Vel. The PMS stars also have proper motions and
RVs consistent with $\gamma^2$~Vel and
Vela~OB2, and main-sequence fitting to stars in the wider Vela OB2
association gives a
distance that coincides
with the distance to $\gamma^2$~Vel and the main-sequence fitting
distance to the early-type stars
immediately surrounding it.  Using PMS isochrones and a handful of
observations of lithium in K-type kinematic members, J09 
claimed an age of 5-10 Myr for the low-mass PMS stars. A Spitzer
survey for circumstellar material around the low-mass association
members reported by Hern\'andez et al. (2008) revealed a very low disc
frequency that may be consistent with this age estimate given the short
disc half-life of $\simeq 3$\,Myr found from Spitzer observations of
many other young clusters. On the basis of a 3--4\,Myr
age estimate for $\gamma^2$~Vel made by de Marco \& Schmutz (1999) and North
et al. (2007), J09 suggested a cluster formation
scenario in which the massive binary formed last, heating and
evaporating the remaining gas, unbinding the cluster and terminating
star formation -- a scenario similar to the gas expulsion model for the
unbinding and expansion of OB associations (e.g. Lada \& Lada 2003;
Goodwin \& Bastian 2006) . However, Eldrige (2009) updated the age of
$\gamma^2$ Vel to $5.5\pm 1$\, Myr using models that take account of
rotation and previous mass
transfer between its binary components, weakening the evidence that it
is younger than the surrounding low-mass stars.

The RVs obtained by J09 had comparatively low
precision and could not resolve the association kinematics. The
observed RV dispersion of
2.5\,km\,s$^{-1}$ was interpreted as mostly due to measurement
uncertainties. Any intrinsic RV dispersion would be unable to explain
the presence of stars over the full 10 degree diameter of Vela OB2 if
they were intially in a much more compact configuration. Instead it was
proposed that $\gamma^2$~Vel and its surrounding low-mass siblings are
a subcluster within a larger star forming region responsible for Vela
OB2.

In this paper we present initial results from GES for 
the Gamma~Vel cluster, focused on the
kinematics of its low-mass members. In Sect.~2 we describe the
spectroscopic observations and the measurement of RVs. 
Included in this is an empirical estimate of the RV precision obtained
by GES. In Sect.~3 we describe how we select a kinematically
unbiased sample of cluster members and in Sect.~4 we present an
analysis of the RV and spatial distribution of these members. In
Sect.~5 we discuss our results in the context of the formation and
evolution of the Gamma~Vel cluster and its relationship with Vela OB2. 

\section{Gaia-ESO survey spectroscopy}

\subsection{Target selection and observations}

\label{observations}

\begin{figure}
\includegraphics[width=84mm]{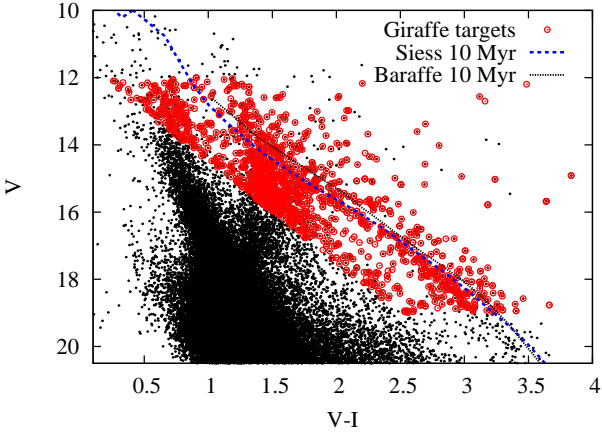}
\caption{A colour-magnitude diagram 
for unflagged objects with good photometry (uncertainties $<0.1$ mag in $V$ and
$V-I$) in a 0.9~deg$^2$ area around $\gamma^2$~Vel (from
Jeffries et al. 2009). Objects with larger red symbols were observed in
the GES. The lines shows theoretical 10\,Myr 
PMS isochrones (from Siess et
al. 2000 and Baraffe et al. 1998, using a colour-$T_{\rm eff}$ relation tuned to match the
Pleiades) at an intrinsic distance modulus of 7.76, and with a reddening and extinction of $E(V-I)=0.055$ and $A_V=0.131$. 
}
\label{vvifull}
\end{figure}

\begin{table}
\caption{A log of the VLT/Flames observations}             
\label{obslog}      
\centering                          
\begin{tabular}{l@{\hspace*{2mm}}l@{\hspace*{2mm}}l@{\hspace*{1mm}}l@{\hspace*{1mm}}l}        
\hline\hline                 
Date & UT & RA, Dec & Exp Time & $N_{\rm targets}$ \\    
     &    & (field centre) & (s) & \\ 
\hline                        
  01 Jan 2012 & 02:39:13  & 08:11:09.2 $-$47:00:03 & $2\times1500$ & 106\\  
  01 Jan 2012 & 03:45:07  & 08:09:10.5 $-$47:02:24 & $2\times1500$ & 100\\
  01 Jan 2012 & 04:53:52  & 08:07:29.3 $-$47:01:15 & $2\times1500$ & 65\\
  01 Jan 2012 & 05:59:32  & 08:11:21.2 $-$47:01:49 & $2\times600$  & 113\\
  02 Jan 2012 & 02:22:15  & 08:10:59.8 $-$47:20:19 & $2\times1500$ & 104\\
  02 Jan 2012 & 03:29:47  & 08:09:46.0 $-$47:20:41 & $2\times1500$ & 111\\
  02 Jan 2012 & 04:33:33  & 08:07:45.9 $-$47:21:07 & $2\times1500$ & 94\\
  02 Jan 2012 & 05:38:39  & 08:09:21.1 $-$47:02:24 & $2\times600$  & 102\\
  03 Jan 2012 & 02:43:43  & 08:10:57.7 $-$47:37:04 & $2\times1500$ & 95\\
  03 Jan 2012 & 03:49:09  & 08:09:22.4 $-$47:37:03 & $2\times1500$ & 94\\
  03 Jan 2012 & 04:55:44  & 08:07:36.1 $-$47:37:04 & $2\times1500$ & 66\\
  03 Jan 2012 & 06:00:44  & 08:07:41.6 $-$47:02:58 & $2\times600$  & 108\\
  15 Feb 2012 & 01:20:04  & 08:11:11.8 $-$47:25:05 & $2\times600$  & 113\\
  15 Feb 2012 & 01:55:33  & 08:09:43.8 $-$47:21:38 & $2\times600$  & 113\\
  15 Feb 2012 & 02:30:39  & 08:07:47.8 $-$47:18:50 & $2\times600$  & 114\\
  15 Feb 2012 & 03:08:03  & 08:10:59.4 $-$47:37:03 & $2\times600$  & 111\\
  15 Feb 2012 & 03:43:01  & 08:09:20.1 $-$47:35:46 & $2\times600$  & 112\\
  15 Feb 2012 & 04:18:28  & 08:07:20.6 $-$47:41:06 & $2\times600$  & 81 \\
\hline                                   
\end{tabular}
\end{table}

The GES strategy for target selection within clusters is described
in detail by Bragaglia et al. (in preparation). Low-mass targets in the
direction of the Gamma~Vel cluster were chosen
primarily by their location in the $V-I$/$V$ and $B-V$/$V$
colour-magnitude diagrams (CMDs) using the 0.9~deg$^2$ photometric survey
of J09. Targets with $12<V<19$~mag, 
corresponding to a mass range of $1.5>M/M_{\odot}>0.2$ (based on an 
intrinsic distance modulus of $7.76\pm0.07$, $E(B-V)=
0.038\pm 0.016$, $A_V = 0.131$ from J09, an assumed age of 10\,Myr and the evolutionary
models of Siess et al. 2000), were selected from the region of the
CMDs surrounding the known spectroscopic and X-ray selected members
in J09. A very wide surrounding margin was considered to ensure that selection
is not biased by these properties. The $V-I/V$ CMD of all stars
is shown in Fig.~\ref{vvifull}, with the objects actually observed (see below) 
indicated. 

The targets were observed with the 
FLAMES fibre-fed spectrographs at the VLT UT-2 (Kueyen) on
the nights of 31 December 2011, 01-02 January 2012, and 14 February 2012.
Both the UVES high-resolution and GIRAFFE intermediate-resolution
spectrographs were used. More than 90 per cent of the spectra were
obtained with GIRAFFE and we deal only with these data here.
GIRAFFE was used in conjunction with the Medusa fibre
system and the HR15n order-sorting filter, which gave spectra with a
resolving power of 17\,000 covering a common wavelength range of 6444--6816\AA.

Targets were grouped according to their $V$ magnitude and configured
for multi-fibre spectroscopy in 18 fields that covered the area of the
photometric survey with significant overlap between fields.  Depending
on target brightness, these fields were observed in ``observation
blocks'' (OBs) of $2\times 600$\,s
or $2\times 1500$\,s. The two exposures were interleaved with a 60\,s
exposure in which 5 dedicated fibres were illuminated by a bright
(compared with the stellar spectra) thorium-argon (ThAr) lamp.  These short
exposures, known as ``simcal'' observations, combined with much longer
day-time ThAr lamp exposures that illuminated all the
instrument fibres, formed the basis of a precise wavelength
calibration. Approximately 20 fibres in each configuration were placed
on blank sky regions and used during the analysis to subtract the sky
contribution from each target spectrum.
The times of observation, central positions, exposure times and number
of targets for each
OB are listed in Table~\ref{obslog}.

A total of 1802 observations of 1242 unique targets were obtained. The overlap between
fields meant that 353 targets were observed twice, 59 were observed 3
times, 23 were observed 4 times and 4 objects were observed on 6
occasions. These multiple observations were used to empirically judge
the precision of the RV measurements. The survey covered 85
per cent of possible targets selected from the photometric survey; the
majority of the unobserved targets have $12<V<16$~mag.

\subsection{Data Reduction}

\label{datared}

Full details of the GES GIRAFFE data reduction will be
given in a forthcoming paper (Lewis et al. in preparation). In
brief, the raw data frames were corrected for a bias level using zero
exposure bias frames and the images were divided by normalised daytime tungsten
lamp exposures to remove pixel-to-pixel sensitivity variations. 
The multiple spectra on each frame were also traced
using the tungsten lamp exposures and then
extracted using the optimal algorithm described by Horne
(1986). This algorithm also yields the 
estimated signal-to-noise ratio (SNR) in the extracted spectral pixels, given
the readout noise and gain of the CCD, and it is this
estimate that is propagated through subsequent analysis steps leading
to the final reported SNR of the spectra. The extracted tungsten lamp
spectra were used to correct the overall shape of the spectrum
and calibrate the individual transmission efficiencies of each fibre.

The wavelength calibration proceeded in two stages. Deep exposures of a
daytime ThAr arc lamp were used to define a
polynomial relationship between extracted spectral pixel and
wavelength, which typically returned an rms difference from the fit of
0.005\AA\ for about 20 lines used in each calibration. This relationship was
subsequently modified by an offset determined from the
positions of the most prominent arc lines in the short ``simcal''
exposures and by
a barycentric correction.  Spectra were rebinned into 0.05\AA\ pixels
using this wavelength solution and sky was subtracted using a median of
the sky spectra corrected for the differing responses of each
fibre.

Several iterations of the data reduction were made including
consistency checks between 
parallel but independent GIRAFFE reduction pipelines operated at the Cambridge
Astronomical Survey Unit (CASU) and at Keele University. The analyses in this
paper are based on the first internal data 
released by CASU to the GES consortium in July 2013 (GESviDR1Final)
and placed in the GES archive at the Wide Field
Astronomy Unit at Edinburgh
University\footnote{http://ges.roe.ac.uk/}.

\subsection{Radial velocities}

\label{rvs}

RVs were determined using two techniques that will be fully described
in a forthcoming paper (Koposov et al. in preparation). A first pass used a standard
cross-correlation method with a grid of synthetic template spectra at a range of
temperatures, metallicities and gravities (Munari et al. 2005) to give
an initial RV estimate. The second pass used a direct
modelling approach that fits each spectrum with a low-order polynomial
multiplied by a template spectrum, with the RV, projected equatorial
velocity ($v \sin i$), temperature, gravity,
metallicity and polynomial coefficients as free parameters.  An
automated emission-line detection procedure excluded emission lines
from the fitting process -- predominantly the H$\alpha$ line in young
members of the Gamma~Vel cluster.  The best-fit was found by
chi-squared minimisation, but the formal RV uncertainties estimated with
this technique were found to be too small, chiefly due
to systematic uncertainties in wavelength calibration.

\subsubsection{Radial velocity precision}

\label{rvprecision}

\begin{figure}
\includegraphics[width=84mm]{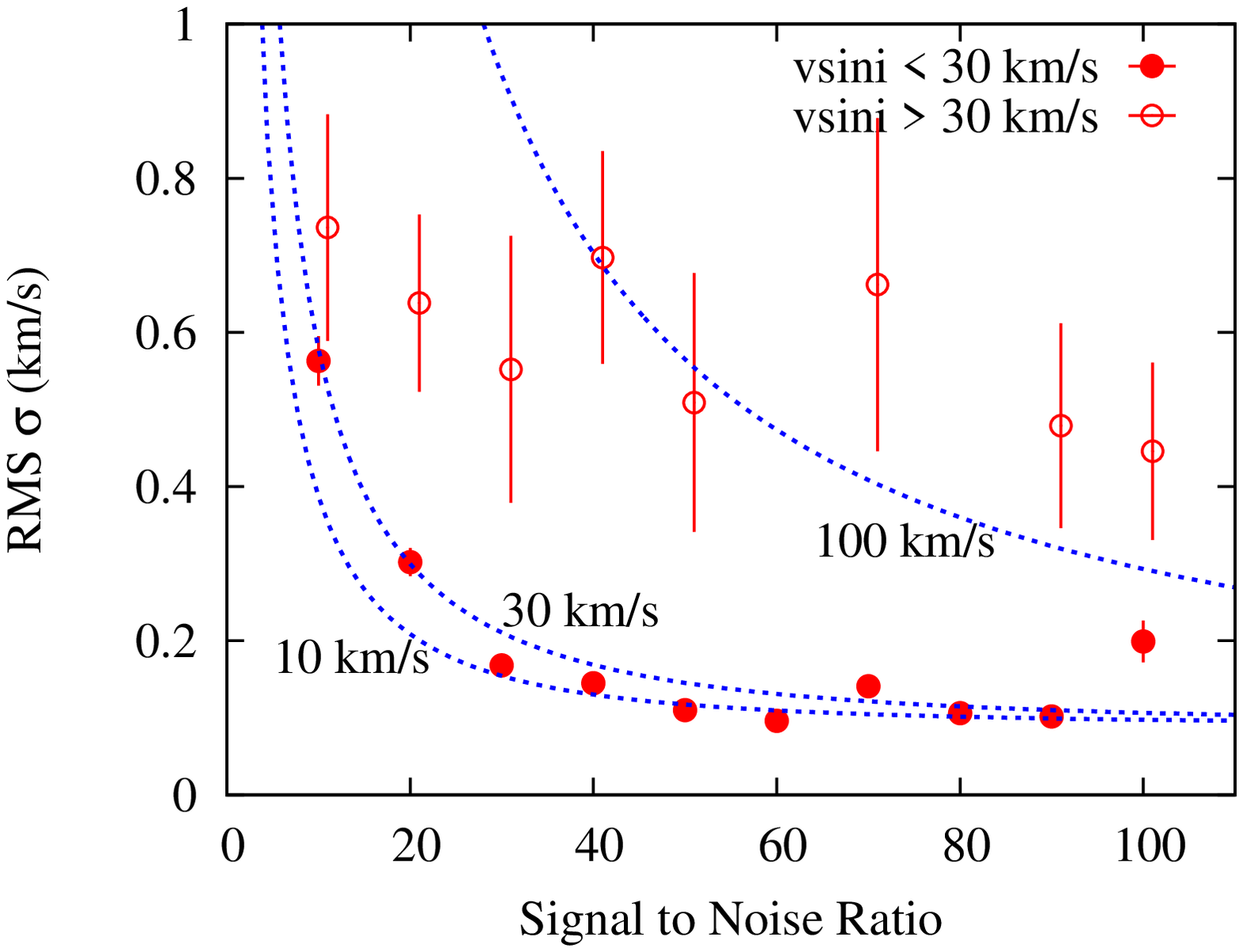}
\includegraphics[width=84mm]{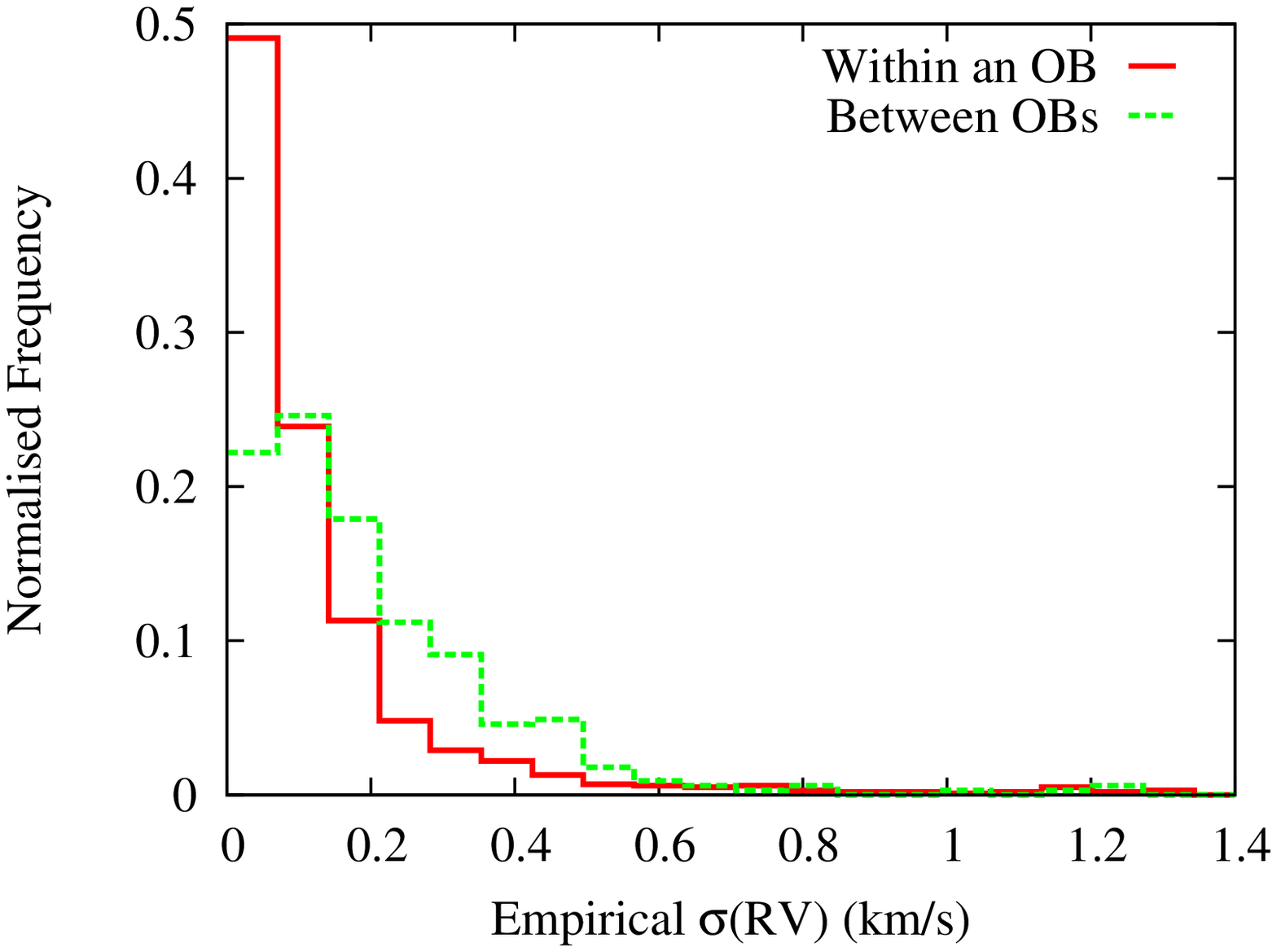}
\caption{The empirically determined RV precision. {\it Upper panel:}
  The rms of the empirically estimated RV uncertainties (see text) from
  pairs of observations within an OB, binned by SNR. A separate set of
  points is calculated for stars that have an estimated $v \sin i > 30$\,km\,s$^{-1}$ to demonstrate their larger empirical
  uncertainties. The lines on the plot are loci determined from
  Equation~\ref{rverr} using the coefficients $A=0.09\pm0.01$,
  $B=3.52\pm0.23$, $C=38\pm8$ for several labelled $v \sin i$
  values. The fit is poorly constrained for large $v\sin i$ and there
  are some indications that the semi-empirical model underestimates the
  uncertainties for such stars at high SNR. 
  {\it Lower panel:} The frequency distribution of empirical RV
  uncertainties determined from repeated observations within an OB and
  from repeated observations from separate OBs. The increase in the
  width of the latter distribution indicates additional uncertainties
  associated with wavelength calibration between OBs. 
}
\label{rverr_fig}
\end{figure}

An accurate assessment of RV precision is key to the dynamical
analysis of clusters. The format of the Gamma~Vel observations lend
themselves to a semi-empirical determination of the RV
uncertainties. Repeated observations of a target in the same and
different fibre configurations allow us to identify and assess various
sources of error.

A series of simulations (see also Jackson \& Jeffries 2010) 
suggest that the RV uncertainty, $\sigma$, has the following 
functional form
\begin{equation}
\sigma^2 =   A^2 + B^2\left[ \frac{1 + (v\sin i/C)^2}{{\rm SNR}}\right]^2
\, , 
\label{rverr}
\end{equation}
where $A$, $B$ and $C$ are constants to be determined, $v\sin i$ is
measured from the spectrum during the chi-squared fitting process
(Koposov et al. in preparation)
and SNR is the median signal-to-noise ratio per pixel of the spectrum
derived during the extraction process. 
This formula allows for the expected decrease in
precision with decreasing SNR and increasing $v \sin i$, but the $A$
term also accounts for any systematic uncertainty associated with the
wavelength calibration. We determine $A$, $B$ and $C$ in two stages. 

(i) We consider 1770 pairs
of exposures of the same object, taken within the same OB and where the
combined spectrum has a SNR$>5$. An estimate of the RV precision in the
combined spectrum is
given by $\sigma = |\Delta$RV$|/\sqrt{2}$, where $\Delta$RV is the change
in RV between the two exposures. We clip out 53 pairs with
$|\Delta$RV$|>2$\,km\,s$^{-1}$. Figure~\ref{rverr_fig} shows the rms value of
$\sigma$ as a function of SNR for the remaining pairs.
$A$, $B$ and
$C$ are estimated by
fitting Equation~\ref{rverr} to a surface of $\sigma$ in the SNR vs $v
\sin i$ plane. 
This yields $A=0.09\pm0.01$, $B=3.52\pm0.23$, $C=38\pm8$ (all in units of
km\,s$^{-1}$). This means that within a single OB, the repeatability
of RV measurements is 90\,m\,s$^{-1}$ for stars with high SNR and small $v \sin i$. Several loci determined using Equation~\ref{rverr} with
these coefficients are shown in Fig.~\ref{rverr_fig} where the
uncertainty is plotted versus SNR for binned data
points with $v\sin i$ above and below 30\,km\,s$^{-1}$. 
Although the model is poorly constrained at high
$v \sin i$ values where there is little data (95 per cent of the
  targets used have $v \sin i < 30$\,km\,s$^{-1}$), it appears that
rotational broadening has little effect below 30\,km\,s$^{-1}$, but
uncertainties increase rapidly thereafter, with some indication that
Equation~\ref{rverr} underestimates the uncertainties for high $v\sin
i$ and SNR by about 30 per cent. 

(ii) The coefficients do not yet account for uncertainties in the
wavelength calibration because an identical calibration is applied to
each set of spectra within an OB.  We assume that $B$ and $C$ are
properties of the observed stars and the fitting process alone and can
be applied to any spectrum. To estimate a value of $A$ that takes
account of wavelength calibration uncertainties we fix $B$ and $C$ and
fit Equation~\ref{rverr} to the RV differences from 329 pairs
of observations\footnote{Where there are $N$ ($>2$) observations of the same
  object we treat these as $N-1$ independent pairs.}
in {\it different} OBs, and where each contributing spectrum
has a SNR$>5$ and the OBs were taken within 3 days of each
other. This latter condition minimises any variations caused by the
motion of unresolved binary systems. Using a model binary
distribution (see Sect.~\ref{rvanalysis}) we estimate that only one per cent
of our targets are expected to be in unresolved binaries that have an RV that
varies by between 0.5\,km\,s$^{-1}$ and 2\,km\,s$^{-1}$ (beyond which they
are clipped in any case) on 3-day timescales. The bottom panel of
Fig.~\ref{rverr_fig} shows that the distribution of empirically
estimated uncertainties becomes broader when considering repeat
measurement from different OBs and we find $A = (0.246 \pm
0.029)$~km\,s$^{-1}$. The rms uncertainty increases from about
0.10\,km\,s$^{-1}$ for repeats within an OB to 0.28\,km\,s$^{-1}$ for
repeats between OBs, though in both cases, a Gaussian
distribution is not a good representation. There is a narrower core and
longer tails, presumably because of stars with low SNR and/or high $v
\sin i$. This is accounted for by our model and is why in our
subsequent modelling we assign an individual RV uncertainty to each
star, based on Equation~\ref{rverr}, rather than using a single average
value.

The $B$ and $C$ coefficients derived here will be
specific to the wavelength range and types of star observed. However, we
anticipate that in most cases the GES RV precision will be dominated by the
$A$ coefficient and so the estimates provided here are likely to be widely applicable
to late-type stars observed in GES with the same instrumental
configuration. Early-type stars with fewer spectral features
and often rapid rotation will have less precise RVs that are determined
by their own particular $B$ and $C$ coefficients.

\subsection{Lithium equivalent widths}

\label{lithium}

The GESviDR1Final spectra include the Li~{\sc i}~6708\AA\ feature that can be used as
a membership indicator (see Sect.~\ref{membership}).
The GES analysis (described in more detail in a forthcoming paper --
Lanzafame et al. in preparation) uses three independent methods for deriving
the equivalent width of this feature in the GIRAFFE spectra (hereafter
referred to as EW(Li)): direct profile integration as implemented in
{\sc splot} as part of the 
{\sc iraf} package\footnote{{\sc iraf} is written and supported by the
  National Optical Astronomy Observatories (NOAO) in Tucson,
  Arizona. NOAO is operated by the Association of Universities for
  Research in Astronomy (AURA), Inc. under cooperative agreement with
  the National Science Foundation}, {\sc daospec} (Stetson \& Pancino
2008) and an {\it ad hoc} procedure
written in {\sc idl} (E. Franciosini private communication). 
The latter automatically
derives EW(Li) and its uncertainty by a direct profile integration
taking into account the star's RV, $v \sin i$, and SNR. 
The analyses
were performed on the summed spectra of each object.

The results obtained by the three different methods were first compared
to check for systematic differences before combining them to produce
the final results. Above 300 m\AA\ the {\sc daospec} results were
discarded because they systematically underestimated EW(Li) in FGK
stars and overestimated it in M stars and fast rotators with respect to
the results obtained using {\sc iraf} (probably because the strong line
is non-Gaussian -- see Pancino \& Stetson 2008). In this range the final EW(Li)
is given as the average between the {\sc iraf} value and {\sc idl} procedure.
Below 300 m\AA\ the results obtained by the three methods were
averaged, eventually discarding one of the three values if it deviated
by more than one standard deviation.
The uncertainty on the final EW(Li) is given conservatively as the
larger of the standard deviation or the average uncertainty
from the independent measurements. The median uncertainty of a detected
Li line is
14\,m\AA\ with a median SNR of 36. Where no significant EW(Li) can
be found, an upper limit is estimated using the approach suggested by
Cayrel (1988). As each of the three EW estimation procedures uses
  an independent approach to establishing a continuum level, the
  quoted uncertainties automatically contain some allowance for
  uncertain continuum placement and this is the dominant source of
  uncertainty even at the median SNR.

The standard GES analysis of EW(Li) also makes an attempt to account, where
necessary, for blending with a nearby weak Fe\,{\sc i} line (Soderblom
et al. 1993) and, in only 14 cases, the presence of a veiling continuum
that is presumably due to accretion and diminishes the measured
EW(Li). We ignore both of these corrections 
in the present analysis, using only the
``raw'', blended EW(Li) that is reported in the GESviDR1Final tables, 
but have confirmed that their inclusion 
would have made no difference to the selection of members
described below.

\section{Membership selection}

\label{membership}

\begin{figure}
\includegraphics[width=84mm]{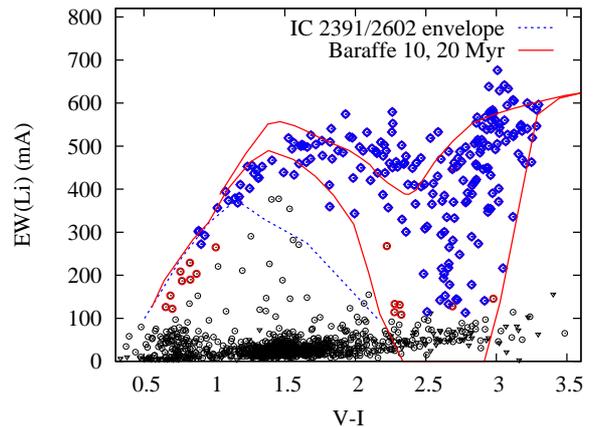}
\caption{The equivalent width of the Li~{\sc i}~6708\AA\ feature
  (EW(Li)) versus
  $V-I$. Objects selected as candidate members are marked
  with blue diamonds; red circles mark objects that are potential
  members based on their photospheric Li content, but fall outside the sequence of members in
  the colour-magnitude diagram (see Fig.~\ref{vvi}). Triangles mark
  upper limits. The red solid lines show theoretical isochrones at
  10\,Myr and 20\,Myr from the models of Baraffe et al. (1998, see
  text), reddened with $E(V-I)=0.055$. 
  The dashed line marks the upper envelope of EW(Li) for IC~2391
  and IC~2602 at ages $\simeq 50$\,Myr. 
}
\label{livi}
\end{figure}

\begin{figure}
\includegraphics[width=84mm]{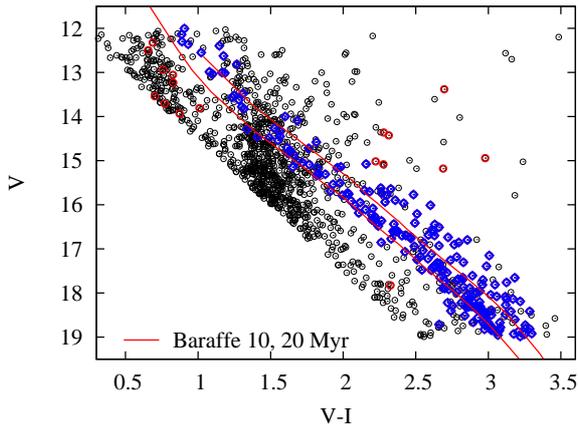}
\caption{A colour-magnitude diagram for the Gamma~Vel
  targets. Blue diamonds show the Li-rich targets we select as members
  (see Fig.~\ref{livi}); red circles are Li-rich targets that do not
  follow the sequence defined by the bulk of cluster members and are
  excluded. Red solid lines show isochrones from the Baraffe et
  al. (1998) evolutionary models (see text), shifted to a distance
  modulus of 7.76~mag, with reddening $E(V-I)=0.055$ and extinction
  $A_V=0.131$ applied.
}
\label{vvi}
\end{figure}

\begin{table*}
\caption{Coordinates, photometry and measurements of RV, EW(Li), $v
  \sin i$ and estimated masses for 208 targets selected as
  Gamma~Vel members. The full table is available in electronic
  format; a sample is shown here.}
\begin{tabular}{ll@{\hspace*{1mm}}llll@{\hspace*{2mm}}l@{\hspace*{2mm}}l@{\hspace*{2mm}}ll@{\hspace*{2mm}}ll@{\hspace*{2mm}}l@{\hspace*{2mm}}l}
\hline\hline
CNAME & RA  & DEC  &  $V$  &  $V-I$ &  SNR  & N & EW(Li) &
$\Delta$EW(Li) & RV & $\sigma$RV & $P_A$ & $v \sin i$  & $M/M_{\odot}$ \\
      & \multicolumn{2}{c}{(J2000)} & & & & & \multicolumn{2}{c}{(m\AA)}
& \multicolumn{2}{c}{(km\,s$^{-1}$)} & & (km\,s$^{-1}$) & \\
\hline
08064077-4736441&	08:06:41&	$-$47:36:44&	13.48&	1.31&	135&	1&	425.8&	1.7&	61.14&	0.25&	$-$1.0&	1.2&	1.23\\
08064390-4731532&	08:06:44&	$-$47:31:53&	17.65&	2.67&	21&	1&	178.2&	14.3&	16.54&	0.31&	0.85&	13.7&	0.45\\
08065007-4732221&	08:06:50&	$-$47:32:22&	18.22&	2.72&	13&	1&	452.7&	22.0&	20.89&	0.37&	0.05&	1.0&	0.43\\
\hline
\end{tabular}
\tablefoot{$N$ is the number of individual observations contributing to
  the mean values quoted. $P_A$ is the probability that the star belong to kinematic
  population A, but is set to -1 if the RV is outside the range 8 to
  26\,km\,s$^{-1}$. Although values of $v \sin i <10$\,km\,s$^{-1}$ are
  reported in the GESviDR1Final tables, we suspect that subsequent
  analysis will likely suggest these are unreliable and we treat them
  as upper limits at 10\,km\,s$^{-1}$ in Sect.~\ref{rotation}.
  Masses are estimated from the $V-I$ colour and
  models of Baraffe et al. (1998) for an assumed age of 10\,Myr.}
\label{tab_members}
\end{table*}

For this paper, our aim is to select a sample of association members
as free from any kinematic bias as possible. The initial
selection of candidate members therefore {\it did not} use the RV results.
Instead we rely on the presence and strength of the Li\,{\sc
  i}~6708\AA\ feature and the position of candidates in
the $V-I/V$ CMD.

Lithium is a well-known age indicator for
young PMS stars because
it is rapidly depleted if the temperature at the base of
the convection zone, or core of a fully convective star, exceeds $\simeq
3\times 10^{6}$\,K  (e.g. Soderblom 2010). 
Theoretical isochrones of Li depletion have been
calculated (e.g. Baraffe et al. 1998; Siess, Dufour \& Forestini 2000),
but are subject to significant uncertainties regarding convective
efficiency and atmospheric opacities. The models match the broad
picture that has emerged from observations of Li abundances in young
clusters, but do not agree with each other and cannot explain the
significant scatter in Li abundance often seen in stars with similar
age and $T_{\rm eff}$. Here we adopt an empirical approach
and use a large EW(Li) as the principal criterion for selecting cluster
members. The timescale for significant Li depletion ranges from $\sim 10-20$\,Myr in
mid-M stars, $\sim 100$\,Myr in K-type stars, to $\sim 1$\,Gyr in
G-type stars. Thus,
the presence of Li will exclude the vast majority of field K- and
M-dwarfs, but will not be as effective at excluding contaminating field
G dwarfs. 

Figure~\ref{livi} shows EW(Li) plotted against
$V-I$ colour  compared with 10 and
20 Myr isochrones calculated from the evolutionary models of Baraffe et
al. (1998 -- with mixing length of 1.0 pressure scale heights). The
model Li abundances are folded through the NLTE curves of growth
described in Jeffries et al. (2003), based on {\sc atlas9} models for
warmer stars ($T_{\rm eff} > 4000$\,K, Kurucz 1993) and from
Zapatero-Osorio et al. (2002) for $T_{\rm eff} \leq 4000$\,K. The
relationship between $T_{\rm eff}$ and $V-I$ is also that used in
Jeffries et al. (2003 -- see their fig.~6) and we redden the models by
$E(V-I) = 0.055$. The 20\,Myr isochrone is a
{\it guide} to the lowest EW(Li) we might expect from Gamma~Vel
cluster members, but must be used with caution given the
sensitivity of such models to convective efficiency and the details of
the atmosphere and curves of growth.  A further empirical locus
marks the
observed upper envelope of EW(Li) in the clusters IC~2391 and IC~2602
(Randich et al. 1997, 2001), that have ages of
about 50\,Myr and which we expect more definitely define a lower
boundary for association members. $E(V-I)$ values of 0.01 and 0.04 were assumed for
these clusters (Patten \& Simon 1996). Caution is still warranted in stars
with $V-I < 1.5$, where any small systematic differences in the EW(Li)
measurements between our work and the literature samples (e.g due to the
continuum definition, or how rapid rotation is dealt with) 
could be comparable to the Li depletion expected between 10 and 50\,Myr.

The larger symbols (both blue diamonds and red circles) in Fig.~\ref{livi} 
show those objects we initially select as Gamma~Vel cluster
members on the basis of EW(Li) and the loci discussed above. A
significant detection of the Li line with EW(Li)\,$>100$\,m\AA\ was taken
as an absolute criterion. Contamination by Li-rich field giants is still possible, but is
ignored because only $\sim 1$ per cent of G/K giants might exhibit
photospheric Li at a level that meets our threshold (Brown et al. 1989). 
The Li-selected targets are
then passed through a further filter in the
$V-I/V$ CMD (shown in Fig.~\ref{vvi}). Here, the purpose is to
exclude Li-rich objects that appear to lie well away from the
locus defined by most cluster members. Figure~\ref{vvi} shows that,
as expected, the presence of Li in G stars is not a reliable indication
of cluster membership and most of these have been excluded by the
CMD filtering. Few stars at cooler spectral types are excluded in the
CMD because field stars at these temperatures are rarely expected to
retain Li at the levels we demanded in Fig.~\ref{livi}.

The definition of these two filters is to some extent arbitrary, our
concern is mainly to avoid contamination by field stars. We note that
it is possible that we have excluded a few genuine cluster
members. These might be objects that are displaced in the CMD by
variability or their photometric uncertainties, 
but our filter should not exclude unresolved binaries.  It
is also possible we have excluded a few objects that have an apparently
weak Li line due to very rapid rotation or,
particularly around $V-I\sim 2.7$, have actually depleted Li beyond the
limits we have accepted. On the basis of the RVs of excluded objects
and the small fraction of objects showing evidence for rapid rotation in
our sample, the numbers of excluded genuine members is unlikely to
amount to more than a few per cent of the accepted sample.

\section{Results and Analysis}

\begin{figure}
\includegraphics[width=84mm]{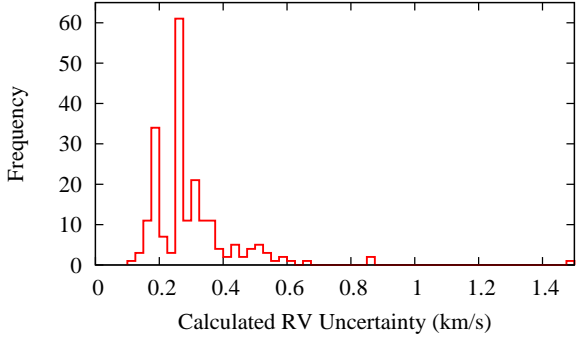}
\caption{The distribution of RV uncertainties for the
  selected Gamma~Vel members, calculated using
  Equation~\ref{rverr}. Where objects were observed in more than one
  OB, the uncertainty in the weighted mean RV is correspondingly smaller.
}
\label{rverrhist}
\end{figure}

All the data for all the targets observed in the Gamma~Vel OBs
listed in Table~\ref{obslog}, including the EW(Li) values, photometry
(from J09) and RVs (including RVs determined from individual exposures
and from OBs) can be obtained from the Gaia-ESO archive hosted by the
Edinburgh University Wide Field Astronomy Unit. Results for the 208
Gamma~Vel members selected in Sect.~\ref{membership} are
reported in Table~\ref{tab_members} (available in electronic format).  For each cluster member we report
the weighted mean RV (weighted by the uncertainties calculated from
Equation~\ref{rverr} for each contributing OB) and its total calculated
uncertainty, the mean rotational broadening and the total 
SNR (across multiple OBs where
appropriate) and the mean EW(Li) and its uncertainty obtained as described
in Sect.~\ref{lithium}. Figure~\ref{rverrhist} shows the
distribution of calculated RV uncertainties for the Gamma~Vel
members. The median uncertainty is 0.26\,km\,s$^{-1}$. The obvious bimodality
arises from the 80 objects with more than one independent observation
and consequently smaller final RV uncertainties.

\subsection{Modelling the radial velocity distribution}

\label{rvanalysis}

\begin{table}
\caption{Results of the maximum likelihood RV modelling.}
\begin{tabular}{ccc}
\hline\hline
     & One component & Two Components \\
\hline
RV$_A$  (km\,s$^{-1}$)  &  $17.71\pm0.14$  (17.72) & $16.73\pm0.09$ (16.72)     \\
$\sigma_A$ (km\,s$^{-1}$) & $1.63\pm 0.13$ (1.62)  & $0.34\pm0.16$ (0.30)    \\
$\sigma_B$ (km\,s$^{-1}$) &    &$1.60\pm0.37$ (1.85)  \\
$\Delta$RV$_{AB}$ (km\,s$^{-1}$) &    & $2.15\pm0.48$ (1.88)  \\
$f_A$                     &    & $0.48\pm0.11$ (0.43)  \\
$\ln L_{\rm max}$          & $-416$   & $-395$    \\
$P_{\rm KS}$                        & 0.006   & 0.994 \\
\hline
\end{tabular}
\tablefoot{Symmetric 68 per cent confidence intervals for one parameter
  of interest, the value at the maximum likelihood fit is given in
  brackets. RV$_A$ is the centre of the first (or only) velocity
  component; $\sigma_A$ and $\sigma_B$ are the {\it intrinsic} 
  velocity dispersions of
  the first and second component; $\Delta$RV$_{AB}$ is the velocity separation
  of the two components; $f_A$ is the fraction of stars belonging to
  the first component; $\ln L_{\rm max}$ is the log likelihood value
  for the best fit and $P_{\rm KS}$ is the probability that the data are drawn from the model as judged by a Kolmogorov-Smirnov test (see text).}
\label{likelyresults}
\end{table}

\begin{figure}
\includegraphics[width=84mm]{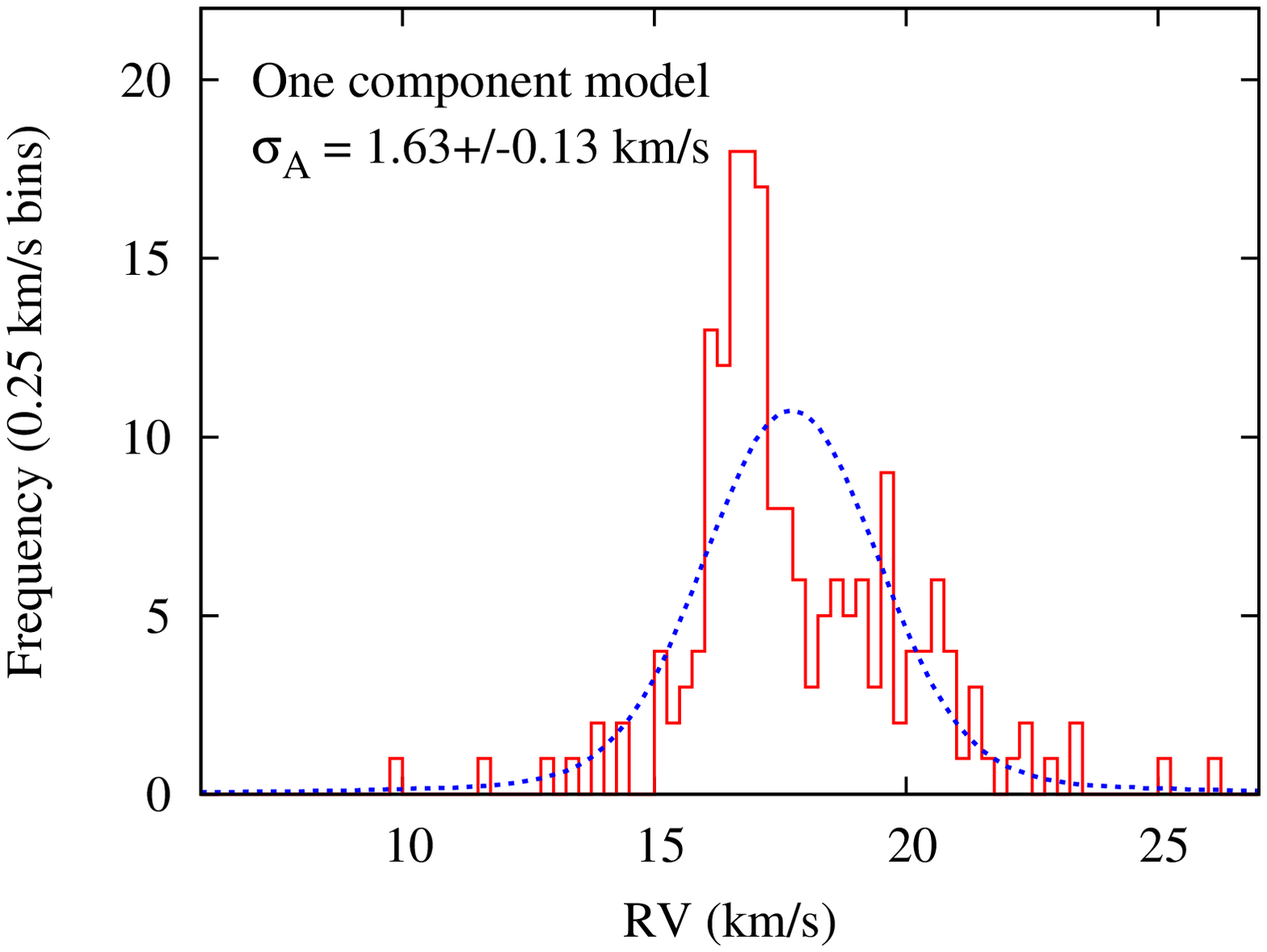}
\includegraphics[width=84mm]{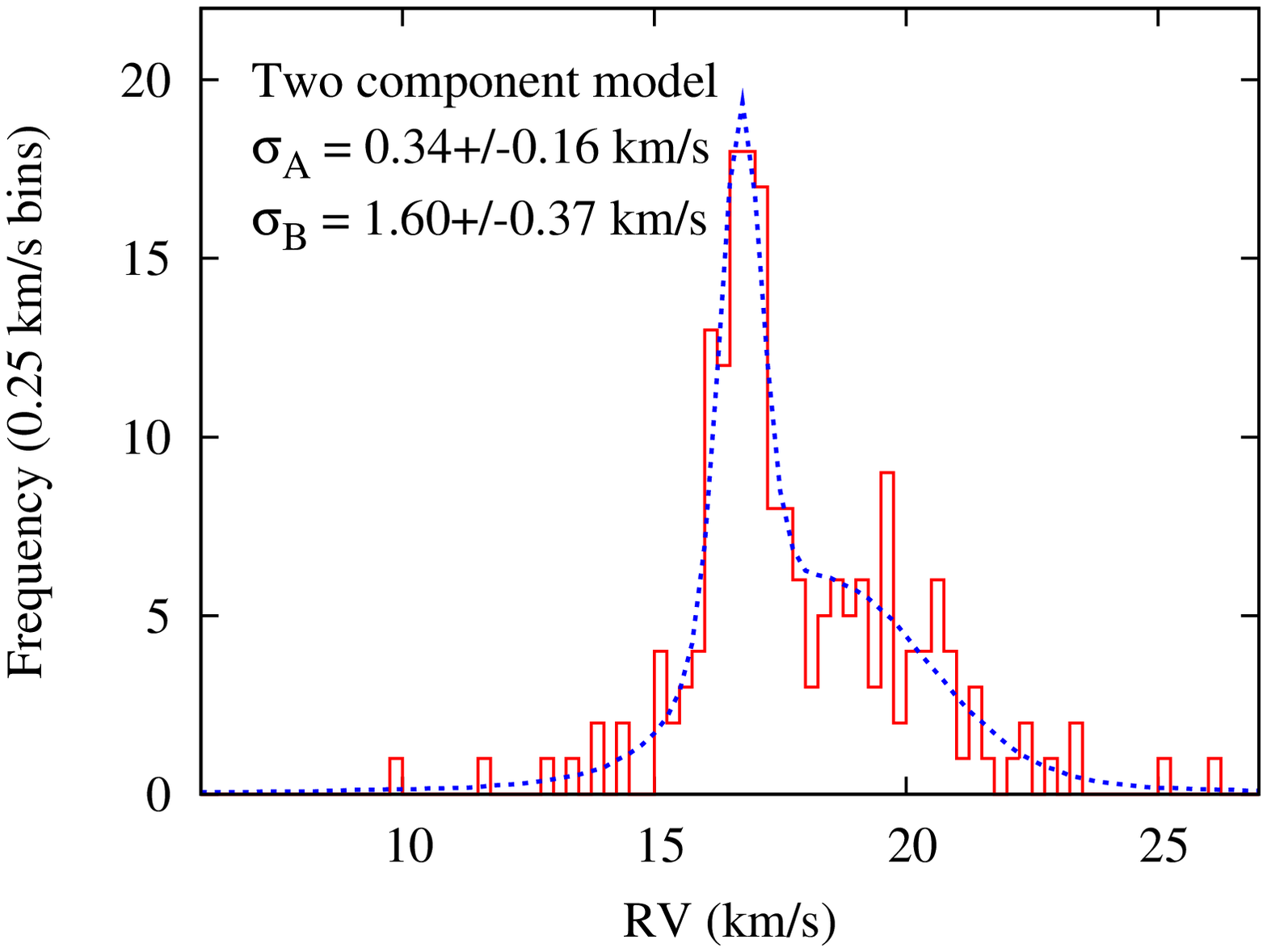}
\caption{A binned RV histogram for the Gamma~Vel members (note
  that fitting was carried out on unbinned data). {\it Upper panel:}
  the best fit for a  model consisting of
  a single Gaussian population with a fraction $f_{\rm bin}=0.46$ of
  unresolved binaries (see text). The fit is poor. {\it Lower panel:}
  the best fit for a population represented by two Gaussian components,
  each with an unresolved binary population.
}
\label{gaus1}
\end{figure}

\begin{figure}
\includegraphics[width=84mm]{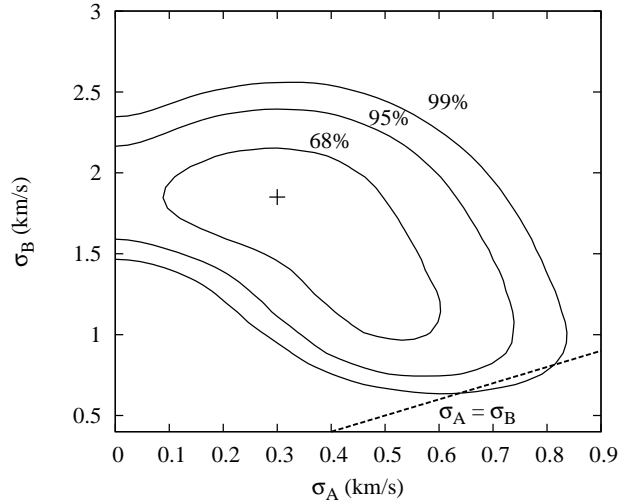}
\caption{The likelihood space for $\sigma_A$ versus $\sigma_B$ for the
  two-Gaussian fit to the RV distribution. Confidence contours are
  shown for two parameters of interest. $\sigma_B > \sigma_A$ with a
  high degree of confidence.  
}
\label{contour}
\end{figure}

A histogram of the mean RVs for stars selected as members is shown in
Fig.~\ref{gaus1} and is modelled using a maximum likelihood
technique. We implicitly assume that our membership selection procedure
has excluded unassociated field stars.
A complication is that some fraction of these objects will be
unresolved binary systems. The procedure we adopt is described in
detail by Cottaar, Meyer \& Parker (2012b) but is summarised here with
some minor differences highlighted.

We assume that the observed RVs are drawn from an intrinsic
distribution that is broadened by measurement uncertanties and the possibility
of binary motion. Single stars and the centres of mass for binaries are
assumed to share the same intrinsic RV distribution. The
likelihood of a star's observed RV, $v_i$, given an intrinsic RV distribution and
the estimated RV uncertainty, $\sigma_i$ is
\begin{equation}
L_i(v_{i}) = (1-f_{\rm bin})\, L_{\rm single}(v_{i},\sigma_i) + f_{\rm bin}\,
L_{\rm binary}(v_i,\sigma_i)\, ,
\label{likelihood}
\end{equation}
where $f_{\rm bin}$ is the fraction of observed objects that are
unresolved binaries; $L_{\rm single}$ is the convolution of a model
intrinsic RV distribution with a Gaussian of dispersion $\sigma_i$; and
$L_{\rm binary}$ is the
equivalent likelihood distribution for binary systems, but is
calculated after convolving the model intrinsic RV distribution with an uncertainty
{\it and} the distribution of velocity offsets expected from a set of randomly
oriented SB1 binary systems, with a specified distribution of orbital periods
and eccentricities.

For binaries, we assume $f_{\rm bin}=0.46$, a lognormal period
distribution, a mean $\log P = 5.03$ (in days) and dispersion 2.28
dex, with a flat mass ratio distribution for $0.1<q<1$ (Raghavan et al. 2010).
For ease of computation we consider only circular orbits; tests using
an eccentricity distribution showed that it has no significant effect on the results.
The binaries are assumed to have a random orientation in space and to
be observed at a random phase of their orbits. Monte Carlo simulations give a
distribution of observed RV offsets for the primary star with respect
to the binary centre of mass. The calculation was
performed separately for each target, assuming a primary mass (given in
Table~\ref{tab_members}) approximated by
interpolating its $V-I$ colour along a 10\,Myr Baraffe et al. (1998)
isochrone and a system mass a factor of $(1+q)$ larger.\footnote{We
  initially performed the analysis using a fixed mass of
  $1\,M_{\odot}$. The best-fitting intrinsic RV distributions and
  parameters differed by much less than the uncertainties in the best-fitting
  parameters presented
  here, indicating that the procedure is quite robust to mass uncertainties.}

Given a model intrinsic RV distribution described by a number of free
parameters (see below), the best-fitting model is found by calculating
the likelihood for each star (from Equation~\ref{likelihood}) and then
maximising the summed $\log$ likelihood for all stars 
by varying the parameters over a
grid of possible values. Uncertainties in a parameter
are calculated from the distribution of maximum $\log$ likelihoods
for that parameter evaluated after optimisation with respect to 
all other model parameters.

\subsubsection{A single Gaussian population}

We begin by considering an intrinsic RV distribution modelled with a
single Gaussian of width $\sigma_A$ and centre RV$_A$. The fit was made
only to data with weighted mean RV between 8 and 26\,km\,s$^{-1}$. There are 18
objects that lie outside this range. All must be
considered candidate binary systems. Only four have multiple
measurements, but these do not show evidence of RV variability at the 1
km\,s$^{-1}$ level. The most likely fit to the remaining 190 objects has
$\sigma_A = 1.63 \pm 0.13$\,km\,s$^{-1}$ and RV$_A = 17.71 \pm
0.14$\,km\,s$^{-1}$. An approximation\footnote{In the maximum
  likelihood fitting, each star has its own RV uncertainty, but we have
  to assume some mean level of uncertainty to broaden the intrinsic RV
  distribution for plotting purposes.} to this model is shown in
Fig.~\ref{gaus1}, where the intrinsic distribution has been
broadened by the mean uncertainty profile (note that this is not the same
as a Gaussian with a dispersion equal to the mean RV uncertainty) and a
fraction ($f_{\rm bin}=0.46$) of the model realisations are broadened
due to binary motion as described earlier. The fit looks poor, but
maximum likelihood fitting does not yield a ``goodness of fit''
parameter. Instead we have compared the cumulative RV distributions of
data and model using a Kolmogorov-Smirnov (KS) test, which rejects the
hypothesis that the data are drawn from the model distribution with
99.4 per cent confidence.

\subsubsection{A two component model}

\label{twocomponent}

The data suggest the addition of a second Gaussian component. We refer
to the two components, or populations, as A and B. With the
binary parameters fixed, this model has 5 free parameters: the
central RV of one population, RV$_A$, the difference in central RV of
the two components $\Delta$RV$_{AB}$, the Gaussian dispersions of the two
components $\sigma_A$ and $\sigma_B$ and the fraction of stars that
belong to the first component $f_A$. It is assumed that the
stars belong to one component or the other so that $f_B = 1- f_A$.
The likelihood of observing a given RV is now
\begin{equation}
L_i = f_A\, L_{A,i} + (1-f_A)\, L_{B,i}\, ,
\label{twocomponentequation}
\end{equation}
where $L_A$ and $L_{B}$ are likelihoods calculated using
Equation~\ref{likelihood}, but with the appropriate intrinsic model RV
distributions for components A and B respectively.

The values of the parameters at the maximum likelihood fit are given in
Table~\ref{likelyresults} along with symmetric 68 per cent
confidence intervals for a single parameter of interest. 
An
approximation to the best-fitting model (calculated using the mean
uncertainty profile) is shown in the lower panel of Fig.~\ref{gaus1}
and consists of roughly equal numbers of stars in each component, one
with a very narrow intrinsic dispersion and the other much broader and
offset by 2.15\,km\,s$^{-1}$.
The maximum log likelihood increases by 21 with this more complex model at the
cost of three additional degrees of freedom. Using the Wilk's theorem
approximation, the two component model is preferred over the single
component model with 99.99 per cent confidence. The two component model
is also preferred according to the Bayesian information criterion. A KS
test yields $P_{\rm KS}=0.994$, suggesting the data and model are
consistent and that searching for more complex structure is unlikely to
yield further significant improvement.

Whilst there are roughly equal numbers of stars belonging to each of
the two components ($f_A = 0.48 \pm 0.11$), their RV dispersions are
quite different. To test whether
correlations between the various parameters might affect this
conclusion Fig.~\ref{contour} shows the maximum likelihood space in the
$\sigma_A$ versus $\sigma_B$ plane. Confidence contours are
calculated according to the usual increments in log likelihood for {\it
  two} parameters of interest. We find there is no strong
correlation and that $\sigma_B > \sigma_A$ with a high degree of
confidence. Similarly, we find that the two components have different RV
centroids with a high degree of confidence; i.e. $\Delta$RV$_{AB} > 0$.

The binary parameters we have used in our models come from a study
  of solar-type field stars. About half our sample has lower masses and
  there is some evidence that the binary frequency is smaller for such
  objects -- perhaps 30 per cent (Duch\^ene \& Kraus 2013). In any
  case, there may also be some difference in binary parameters
  associated with birth environment, so it is prudent to investigate
  the sensitivity of our results to this. We repeated the analysis with
  a binary frequency of 30 per cent. The main difference is that
  $\sigma_B$ increases from 1.60\,km\,s$^{-1}$ to 1.89\,km\,s$^{-1}$, a
  change that is less than the original uncertainty estimate. All other
  parameters change by much less than the uncertainties listed in
  Table~~\ref{likelyresults}, so we conclude that the results are quite
  insensitive to plausible uncertainties in the binary parameters.

\subsection{Two populations in the Gamma Vel cluster?}

\label{twopop}
\begin{figure}
\includegraphics[width=84mm]{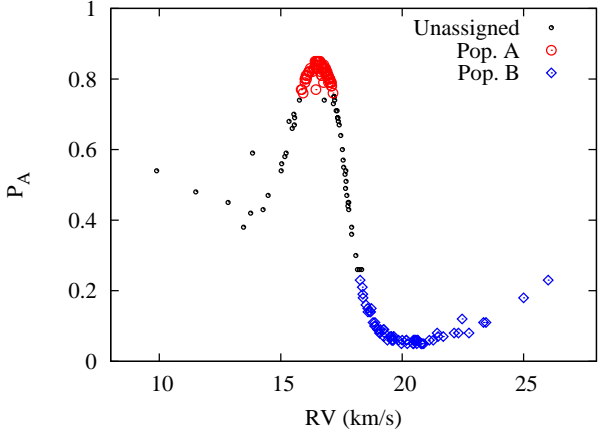}
\caption{
The probability that each Gamma~Vel member belongs to population
A in the two-component kinematic model described in
Sect.~\ref{twocomponent}. Relatively secure members of population A
($P_A > 0.75$)
and B ($P_A<0.25$) and objects that cannot be confidently assigned to either
population are identified with different symbols in this and subsequent plots.
}
\label{probfig}
\end{figure}
Having established (at least) two kinematic components in the RV distribution of
Gamma~Vel members, we can ask whether there are any other
properties that might distinguish these populations. We consider the
colour-magnitude diagram, the spatial distribution, the lithium
depletion, rotational broadening and proper motions. 
In each case it is necessary
to distinguish between members of the two populations, but this can
only be done in a probabilistic way. For our two-component
model we can calculate, for each star, the probability that it
belongs to either populations A or B as
\begin{equation}
P_{A,i} = f_A \frac{L_{A,i}}{L_{i}}\ \ \ \ \ \ P_{B,i} =
(1-f_A)\frac{L_{B,i}}{L_i}\, ,
\end{equation}
where the terms are as defined in Equation~\ref{twocomponentequation}. This could
be calculated for the maximum likelihood model, but we obtain a more accurate
estimate by integrating this probability over the full 5-D
parameter space and then $P_{A,i}$ is found from the expectation value of the total probability
distribution function. $P_{B,i}$ is given by $1-P_{A,i}$.
The results of this calculation are illustrated in Fig.~\ref{probfig}
and listed in
Table~\ref{tab_members} for each star. Stars outside the range
$8<$RV$<26$\,km\,s$^{-1}$ are not considered and given $P_A = -1$. 
$P_A$ depends mainly on the RV
of the star but also to a lesser extent on the RV uncertainty. To test for differences in
properties between the two populations we adopt two approaches. (i)
Where appropriate, we use $P_A$
and $P_B$ as statistical weights in determining mean properties. (ii)
For plotting purposes and also as a way of more carefully cleaning the
samples of contamination, we divide the sample
into a population of 73 objects having $P_A>0.75$, which we will call population A,
though we understand it will still have $\simeq 19$ per
cent contamination by stars in the other population, and 66 stars with $P_A<0.25$
that we call population B, but will have $\simeq
9$ per cent contamination from the other population (see
Fig.~\ref{probfig}). The remaining 69
objects cannot be assigned to either population with great confidence and
are plotted with different symbols.

\subsubsection{The colour-magnitude diagram}
\begin{figure}
\includegraphics[width=84mm]{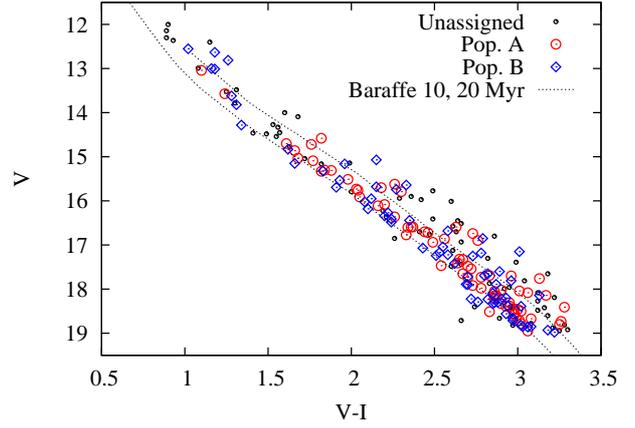}
\caption{
A colour-magnitude diagram for the two kinematic populations around
$\gamma^{2}$~Vel. The populations are distinguished here by a
simple probability
criterion (see Sect.~\ref{twopop}) 
into population A (with a narrow RV distribution) or 
population B (with a broader RV distribution). The other cluster
members cannot
be assigned to either population with great confidence.
}
\label{cmdab}
\end{figure}
Figure~\ref{cmdab} shows the $V-I/V$ CMD for members, coded
for populations A or B according to the probabilistic criterion
described above. The unassigned objects and objects with RV outside
the range $8<$RV$<26$\,km\,s$^{-1}$ are also shown for
completeness. 
We fit the distribution of points with a quadratic in
$V-I$ for $V-I>1.5$~mag, as the few points bluer than this define a kink in
the diagram that is poorly reproduced by a polynomial. Fixing the
linear and quadratic terms, we then fit populations A and B separately,
allowing a constant offset. If the division is carried out as in
Fig.~\ref{cmdab}, the difference in offsets for the two populations 
is a marginal $0.06 \pm 0.07$\,mag
in the sense that population A is brighter. If instead we perform fits weighted
according to the probability of membership of populations A or B, this
difference becomes $0.02 \pm 0.03$\,mag. 
At an age of about 10\,Myr the latter result and the Baraffe et
al. (1998) isochrones plotted in Fig.~\ref{cmdab}
imply that population A is
younger than population B by about $0.4\pm 0.6$\,Myr if they are at the same
distance and have the same unresolved binary frequency. 
Alternatively, if they both have the same age, then
population A is closer by $4\pm 5$\,pc. 

A KS test of the mass distributions of the two
populations resolved according to the probability criteria,
reveals no significant difference ($P_{\rm KS}=0.84$). There are 8
higher mass objects in population B with $V-I<1.5$, and only 2 in
population A. This difference is not regarded as significant by a KS
test though it is marginally significant (at the 95 per cent level)
using a two-tailed Fisher test.

\subsubsection{The spatial distribution}

\label{spatialab}

\begin{figure*}
\includegraphics[angle=0, width=150mm]{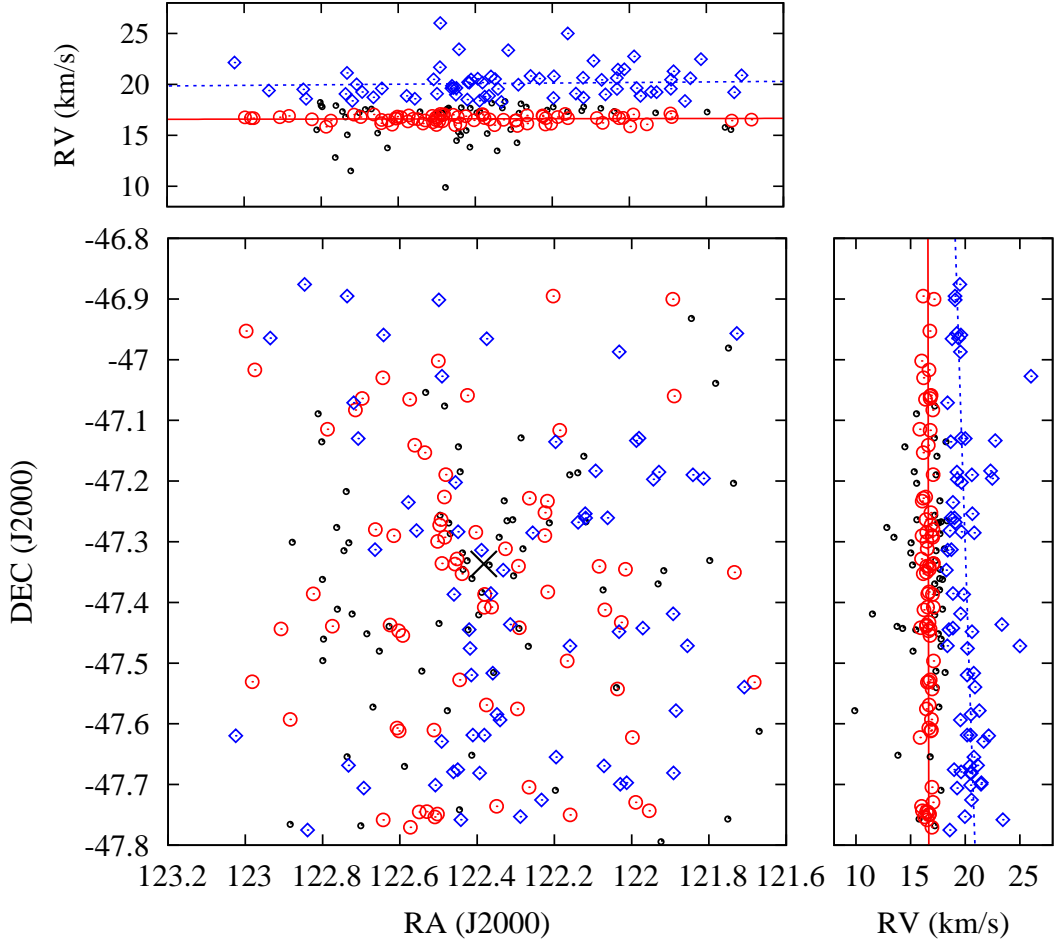}
\caption{The spatial distribution of populations A and B, defined
  according to the probability criterion described in
  Sect.~\ref{twopop}.  In the main panel these populations and the set
  of cluster members that cannot be confidently assigned to either
  are shown. Symbols are as defined in Fig.~\ref{cmdab}.
The cross marks the position of $\gamma^{2}$~Vel. In the
  top and side panels we show the distribution of RV with right
  ascension and declination respectively. The solid (red) lines and the
  dashed (blue) lines show best-fitting linear relationships for
  populations A and B respectively.
}
\label{spatial}
\end{figure*}
\begin{figure}
\includegraphics[width=84mm]{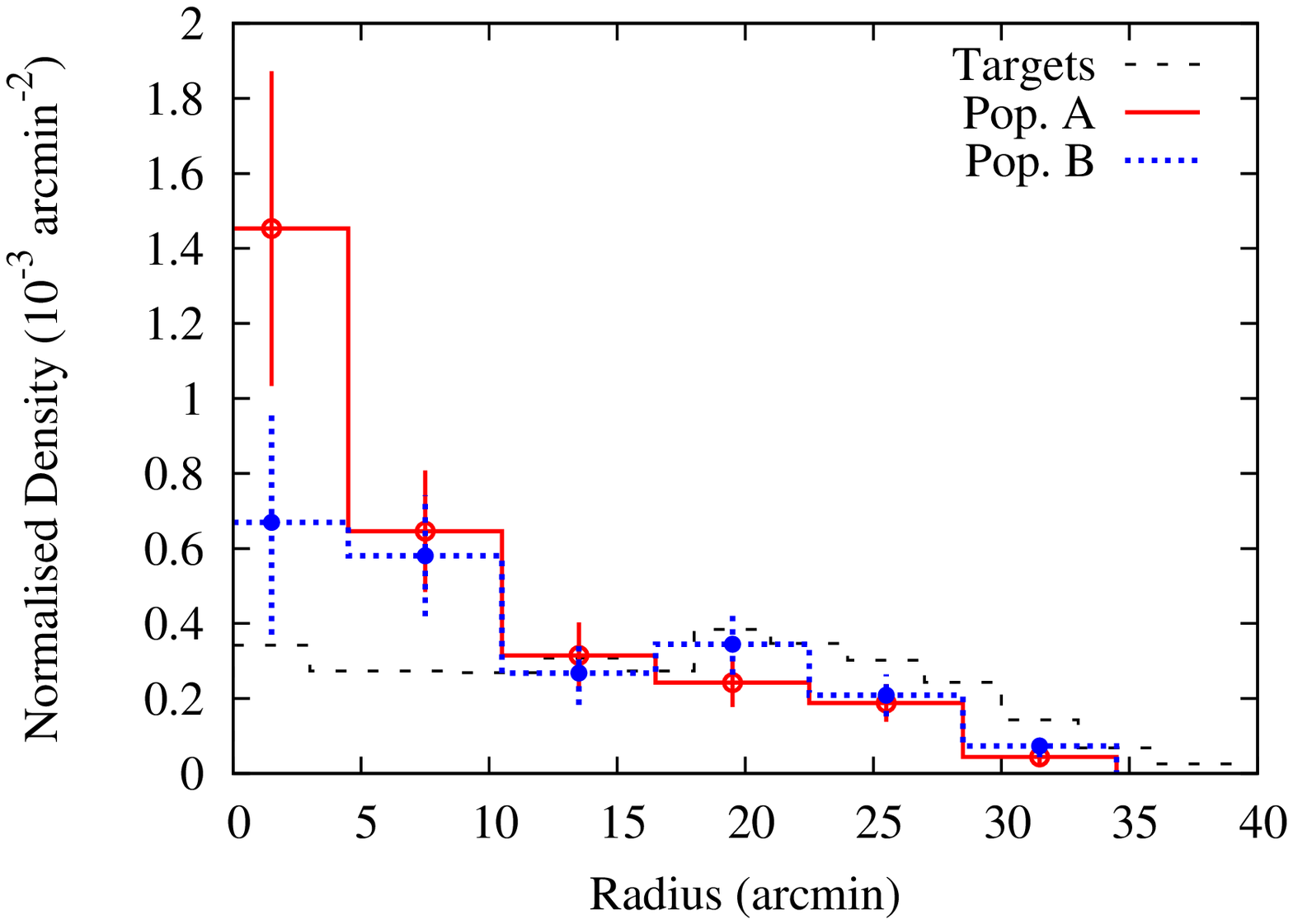}
\caption{
The projected space density of members of populations A and B, along
with that of all the (non-members of the Gamma~Vel cluster) targets
observed. 
Population A is more centrally concentrated than population
B, though the significance of this is low (see Sect.~\ref{spatialab}).
}
\label{spacedensity}
\end{figure}
Figure~\ref{spatial} examines the projected spatial distributions of
the two populations, with different symbols corresponding to the
populations as in Fig.~\ref{cmdab}. The position of $\gamma^2$~Vel is
marked. The most obvious difference between the populations is their
centroid. Population A has a centroid of (RA$=122.413\pm0.034$, Dec$=-47.378\pm 0.027$), whilst the centroid of population B is at (RA$=122.305\pm
0.038$, Dec$=-47.381\pm 0.032$). The distributions of the two populations
are different in RA according to a KS test ($P_{\rm KS}=0.03$), but not
in Dec ($P_{\rm KS}=0.64$).
Accompanying this shift there appears to
be an increase in RV of $\sim 2$\,km\,s$^{-1}$ for population B towards
the south-west. 
A weighted linear fit of RV versus position for the population B objects 
gives a slope of $-0.28\pm 0.60$\,km\,s$^{-1}$\,deg$^{-1}$
in the RA direction and  $-1.80 \pm 0.68$\,km\,s$^{-1}$\,deg$^{-1}$ in
the Dec direction. No drifts are apparent for population A. The best
fitting lines are shown in Fig~\ref{spatial}.

There is a hint that population A
is more centrally concentrated than population B. The centroid of
population A is consistent with the position of $\gamma^2$~Vel, whilst
the centroid of population B is separated from $\gamma^2$~Vel by $4.1
\pm 2.3$ arc minutes. A KS
test of the distributions of radial distance from $\gamma^{2}$~Vel
reveals only a marginal difference with $P_{\rm KS}=0.15$. A central
concentration in either population might be due to biases in the target
selection. To check this we examined the spatial distribution of
targeted objects {\em not} considered to be Gamma~Vel members in
Sect.~\ref{membership}, because these should be subject to
  a similar spatial bias in target selection, but should have an
  approximately uniform intrinsic spatial distribution. The normalised distributions of spatial
density with radius for populations A, B and the targeted non-members
are shown in Fig.~\ref{spacedensity}. Population A is more centrally
concentrated than the target population, with a significantly different
radial distribution ($P_{\rm KS} < 10^{-5}$). Population B is also
more centrally concentrated than the targets, but less so than
population A ($P_{\rm KS}=0.02$).
Given that it is still contaminated with $\simeq 6$ stars from population
A, it is conceivable that population B has a quite similar spatial
distribution to the general field population and that population A,
which has $\sim 13$ contaminants from population B, is even more
centrally concentrated. If we use $P_{A}$ and $P_{B}$ to weight the
contributions from all stars, we find the mean radial distances from
$\gamma^2$~Vel are $15.2 \pm 0.6$, $18.8 \pm 1.0$ and $21.8 \pm 0.2$
arcminutes for populations A, B and the targeted non-members
respectively. 

In summary then, Population A has a centroid consistent with the
position of $\gamma^2$~Vel, is definitely more centrally
concentrated than the observed targets and probably more centrally
concentrated than population B. The centroid of Population B is
offset from population A by $4.4\pm 3.0$ arcminutes and has a spatial distribution
that is marginally consistent with either population A or the observed
targets. Population B shows significant evidence for a spatial gradient
in RV in the
declination direction.

\subsubsection{Lithium depletion}

\begin{figure}
\includegraphics[width=84mm]{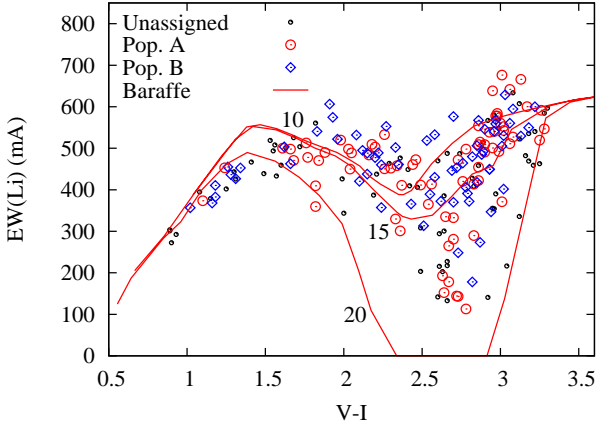}
\caption{
The equivalent width of the Li~{\sc i}~6708\AA\ line versus colour, with different symbols corresponding to the
populations as in Fig.~\ref{cmdab}. The solid lines are isochrones
calculated according to the evolutionary models of Baraffe et
al. (1998, with mixing length of 1.0 pressure scale heights)
at 10, 15 and 20\,Myr, transformed into the observational plane as described in
Sect.~\ref{membership}. 
}
\label{lithiumab}
\end{figure}

Figure~\ref{lithiumab}  shows the EW(Li) versus colour plot, with the
Gamma~Vel members separated by population as described in
Sect.~\ref{twopop}. There is a suggestion in this plot that
population A is more Li-depleted than population B, especially in the
range $2.5<V-I<2.8$ where age-dependent Li-depletion is expected to be
strongest.

The statistics bear this conclusion out. The weighted mean EW(Li) for stars
with $2.5<V-I<2.8$ is $269\pm 21$\,m\AA\ for population A and $433 \pm
19$\,m\AA\ for population B. If we use the whole sample, but
apply a further weighting according to the probability of membership
of populations A or B, the results for the mean EW(Li) in this colour range
are $303 \pm 17$\,m\AA\ for population A and $413\pm 15$\,m\AA, a
smaller difference, but still significant at the 5-sigma level. Hence
this result is quite robust to the details of 
how membership of the two populations is assigned.

The difference in Li-depletion implies a difference in age. Population A
would need to be older (on average) by an amount which is dependent on the
the way that EW(Li) is translated into
a Li abundance. Because the Li~{\sc i}\,6708\AA\ line is in the saturated part
of the curve of growth, a small change in EW(Li) corresponds to a large
change in Li abundance (Palla et al. 2007), but Li-depletion is also
very rapid at these masses, so a large change in abundance is expected
in a small amount of time. Adopting the models of Baraffe et
al. (1998) and the isochrones shown in Fig.~\ref{lithiumab}, which at
least match the colour of dip in Li abundance quite well, we see that
rapid depletion commences at $\sim 10$\,Myr and the small EW(Li)
difference we have found could correspond to only $\sim 1$-2\,Myr in mean
age. The absolute age at which rapid Li depletion commences is model
dependent and begins earlier (but at bluer colours) for models with
higher convective efficiencies, but the implied age difference of $\sim
1$-2\,Myr is reasonably robust to choice of model. That there are
examples of Li-depleted and undepleted stars in both populations with
$2.5 < V-I<3$ might suggest that age spreads are larger than any age
difference. Alternatively it might simply reflect that the samples
identified in Fig.~\ref{lithiumab} are still 
cross-contaminated by the other population. 

Another possibility is that the differences in photospheric Li are due
to composition rather than age differences. Li depletion is
predicted to be very sensitive to interior opacity and if Population A
were even only 0.1 dex more metal-rich than Population~B this might explain
the observed difference (e.g. Piau \& Turck-Chi\`eze 2002). A detailed
compositional analysis of the Gamma~Vel cluster GES UVES data is
underway (Spina et al. in preparation). A detailed discussion of the Li
abundances and comparison with models is deferred to Franciosini et
al. (in preparation); the important point here is that there {\it is} a
difference between populations A and B.

\subsubsection{Rotation rates}

\label{rotation}
\begin{figure}
\includegraphics[width=84mm]{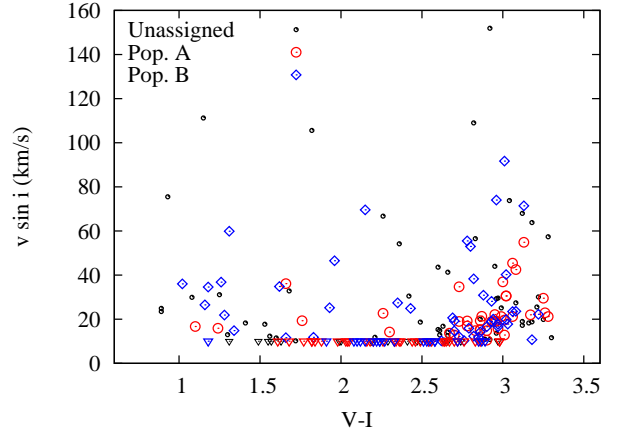}
\caption{Projected rotational velocities as a function of colour.
The different symbols correspond to the various populations
  as described in  Fig.~\ref{cmdab}. Objects with $v\sin i <
  10$\,km\,s$^{-1}$ are marked as upper limits (triangles).}
\label{vsiniab}
\end{figure}

Figure~\ref{vsiniab} shows the projected equatorial velocities of the
sample as a function of colour. Values of $v \sin i < 10$\,km\,s$^{-1}$
are unlikely to be accurate, given the resolution of the spectra, and
are regarded as limits of $<10$\,km\,s$^{-1}$.
There is a hint that members of population B are more rapidly rotating,
on average, especially for $V-I<2.5$~mag, but the numbers are small. 
A two-tailed KS test of the cumulative $v \sin
i$ distributions reveals a marginal difference ($P_{\rm KS} = 0.12$).
Even if the two populations had different ages, it is unlikely that
this comparison would be diagnostic, because the timescale of rotation spindown
with age is too slow ($\sim 50$\,Myr) at the masses of the sample we consider.
Any differences are more likely to reflect different birth conditions.

\subsubsection{Proper motions}
\begin{figure}
\includegraphics[width=84mm]{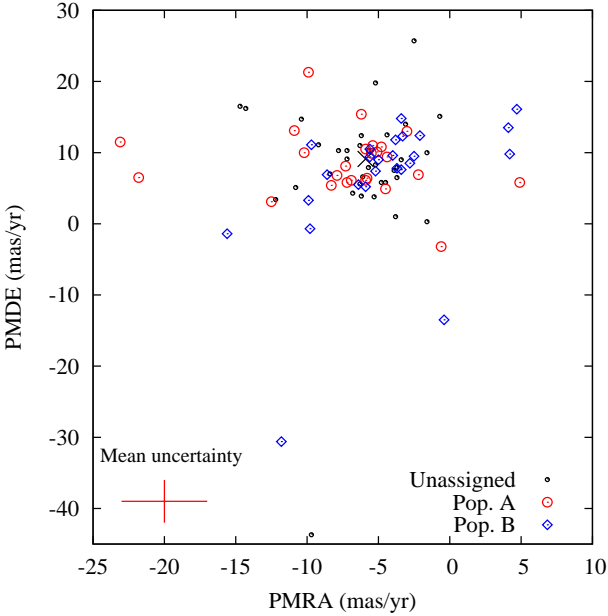}
\caption{UCAC4 proper motions (in milli-arcseconds per year)
  for our selected Gamma~Vel
  members. The different symbols correspond to the various populations
  as described in  Fig.~\ref{cmdab}. The proper motion of $\gamma^2$~Vel
  itself is shown with a cross. There are 6 objects that lie beyond the
  borders of the plot. These and the other very discrepant objects show evidence of
  binarity or lie very close to $\gamma^2$~Vel and are likely to have
  unreliable proper motions.
}
\label{pmab}
\end{figure}

The catalogue of Gamma~Vel members was matched with the Fourth US
Naval Observatory CCD Astrograph Catalog (UCAC4, Zacharias et
al. 2013). There were 92 matches
within 1 arcsecond, most members brighter than $V\sim 17$ had a match.
Figure~\ref{pmab} shows the vector point diagram with population A and
B indicated. There are a number of very discrepant points. We have
examined these stars individually and there are reasons to be
suspicious about all of them; either they appear as blended, unresolved
binaries on photographic plates or they are very close to
$\gamma^2$~Vel itself. Clipping the sample between $-20 <$ PM(RA) $<5$
mas/yr and $-10<$ PM(Dec) $<25$ mas/yr we obtain mean proper
motions of ($-5.9\pm0.8$, $+8.6\pm 1.0$) and ($-4.5\pm 1.0$, $+8.7 \pm
0.9$) mas/yr for populations A and B respectively. The standard
deviations for populations A and B are both $\simeq 4.5$\,mas/yr in 
each coordinate, corresponding 
to $\simeq 7$\,km\,s$^{-1}$ at the
distance of $\gamma^2$~Vel. This is only slightly larger than the respective 
mean proper motion uncertainties and  incapable of resolving
differences in tangential velocity dispersion comparable to that seen
in the RV dispersions. 
If the proper motions are weighted by the probability of population membership,
the mean proper motions of populations A and B are ($-5.9 \pm 0.8$,
$+8.5 \pm 1.0$) and ($-4.6 \pm 1.0$, $+8.7 \pm 0.9$) mas/yr respectively.

These mean proper motions are similar to each other and also similar to
the proper motion of $\gamma^2$~Vel ($-5.9 \pm 0.4$, $+9.9 \pm 0.4$ mas/yr;
from Hipparcos) and the mean proper motion of early type stars in
Vela~OB2 ($-6.6 \pm 1.3$, $+8.1 \pm 1.4$ mas/yr; Tycho proper motions
compiled by J09).

\section{Discussion}

The RV distribution of the young, low-mass stars surrounding
$\gamma^2$~Vel clearly exhibits structure
that has a bearing on the current
dynamical state and the star formation history of the region. The evidence
at hand can be summarised as follows.

\begin{itemize}

\item The RV distribution must be modelled with more than one
  component. A good fit is obtained by dividing the young stars 
 into two, roughly equal, kinematic populations A and B, one with a narrow
 intrinsic dispersion ($\sigma_A = 0.34 \pm 0.16$\,km\,s$^{-1}$) and
 the other much broader ($\sigma_B = 1.60\pm 0.37$\,km\,s$^{-1}$). The
 populations are significantly offset from each other by $2.15 \pm 0.48$\,km\,s$^{-1}$. 

\item Pozzo et al. (2000) and J09 showed that the young, 
  low-mass stars have a spatial density
  distribution that is concentrated towards $\gamma^2$~Vel. Here, we
  have shown that this concentration is mostly evident in population
  A, which has a centroid consistent with the position of
  $\gamma^2$~Vel. 
  Population B appears less spatially concentrated and may be
  consistent with a uniform distribution across our surveyed area. The
  lack of a clean kinematic separation between the populations prevents us being
  more definitive.

\item The two populations are not clearly distinguished from each other in the
  $V-I/V$ CMD or by their rotational
  properties. Their proper motions are similar and consistent with the
  proper motions of $\gamma^2$~Vel and the wider Vela~OB2 association
  that extends well beyond the spatial extent of the GES data.

\item
The mid M-type objects in Population A do show significantly more Li
  depletion than those in Population B. This difference is robust to
  how the populations are separated kinematically and suggests that
  population A could be older on average by perhaps 1-2\,Myr. The
  presence of stars that appear undepleted and depleted in both
  populations implies that their age distributions may overlap, though this
  latter conclusion is again weakened by the lack of clear kinematic
  separation of the populations.

\end{itemize} 

We can hypothesize a number of scenarios for the origin of
these two populations: (i) $\gamma^2$~Vel formed in an originally more massive,
gas-rich cluster. Population A is the bound remnant of the cluster
after gas expulsion and population B is an unbound halo of escaping
stars. We refer to this as the ``core-halo scenario''. 
(ii) $\gamma^2$~Vel
formed in an isolated way but has attracted a retinue of low-mass
stars from the dispersed Vela OB2 association. Population A would be
the ``captured'' stars whilst population B is the dispersed low-mass
component of Vela OB2. We refer to this as the ``captured cluster''
scenario. (iii) $\gamma^2$~Vel formed in a denser sub-clustering of the
Vela OB2 association. At least part of this subcluster survived any gas
expulsion to form population A, while population B consists of the dispersed
low-mass component of the Vela OB2 association. We refer to this as the
``cluster plus association'' scenario. We consider each of these possibilities in turn
with reference to the evidence listed above.
    
\subsection{The core-halo scenario}

It would be very rare for a star as massive as $\gamma^2$~Vel (with
initial mass $\simeq 35\,M_{\odot} + 31.5\,M_{\odot}$; Eldridge 2009) to
form in a cluster with total mass $\sim 100\,M_{\odot}$. The cluster
would be the most distant outlier from the relationship between cluster
mass and mass of the most massive star proposed by Weidner, Kroupa \&
Bonnell (2010), where instead the expected {\it initial} 
cluster mass would be $\sim
1000\,M_{\odot}$. Gas expulsion might solve this problem (and is the
reason that Weidner et al. excluded $\gamma^2$~Vel from their
sample). The fraction of stars lost after gas expulsion depends on the
(local) star forming efficiency, ratio of half-mass radius to tidal
radius, how quickly the gas is lost and the initial clumpiness of the
gas and stars (e.g. Gieles 2010; Bastian 2011). Many simulations show
(e.g. Baumgardt \& Kroupa 2007; Fellhauer, Wilkinson \& Kroupa 2009;
Moeckel \& Bate 2011) that gas removal on less than a dynamical
timescale or a low star formation efficiency can lead to the loss of
most of the original cluster, leaving a bound remnant (Population A),
that may be orders of magnitude less dense than the original embedded
cluster, surrounded by an expanding halo (Population B).

The dynamics of this scenario can be checked using the velocity
dispersions of the two populations.
Clusters in virial equilibrium have RV dispersions ($\sigma_v$) that are related
to their total dynamical masses ($M_{\rm dyn}$) and projected half mass
radii ($r_{m}$) by
\begin{equation}
M_{\rm dyn} = \eta \frac{\sigma_{v}^{2} r_{m}}{G}\, ,
\label{virial}
\end{equation}
where $\eta$ is a numerical constant related to the density profile and
is approximately 10 for a Plummer density distribution (Spitzer 1987; Portegies Zwart, McMillan \& Gieles
2010). 

If we sum up
the Gamma~Vel members in our survey,
weighted according to their probability of belonging to population A,
and assuming that those that lie outside $8<RV<26$\,km\,s$^{-1}$ have
$P_A = f_A = 0.48$, then the total mass of population A (among the GES
survey objects) is 58$\,M_{\odot}$. This mass
estimate should be increased by a factor 1/0.85 to reflect that
GES spectroscopy was obtained for 85 per cent of cluster candidates and
by another factor of 1.25 to account for an assumed binary fraction of
0.46 with a flat mass ratio distribution between 0.1 and 1 (see
Sect.~\ref{rvanalysis}). If we further assume that our survey covers the full extent of
population A and that it is centred on $\gamma^2$~Vel, 
then the half mass radius of this distributed population is $r_m = 0.225$ degrees,
equivalent to 1.37\,pc at the distance of the cluster. 

To these low-mass stars 
we can add the more massive stars that are not included in GES.
These are obtained from the Tycho catalogue and filtered on
proper motion and their position in the $V$ versus $B-V$ CMD to obtain
a list of secure association members (see
J09). We assign masses to these based on their $V$ magnitudes and 
a $Z=0.02$ 10\,Myr isochrone from the Siess et al. (2000) models. 
There are 13 stars which are likely association members and within the
area surveyed by GES, with a total mass of 32\,$M_{\odot}$. However, we
do not know if these belong to populations A or B, so assign a fraction
$f_A$ of the mass to population A. 
The total mass of distributed stars in population A
is therefore $(58\times 1.25/0.85) + (32\times0.48) = 101\,M_{\odot}$,
where we neglect the contribution of any stars below 0.2\,$M_{\odot}$.
We also neglect the contribution from any residual gas. The low and
uniform extinction towards the early type stars of the Gamma Vel cluster
suggests this is justified. 

The
central concentration of population~A and the positional coincidence
between its centroid and the position of $\gamma^2$~Vel, suggest that
$\gamma$~Vel belongs to population A. If so, the total mass of that
system is 39$\,M_{\odot}$ (for $\gamma^2$~Vel; de Marco and Schmutz
1999) and a further 12\,$M_{\odot}$ for its proper motion companion
$\gamma^1$~Vel, which is a resolved multiple with components for which we 
estimate masses of 7\,$M_{\odot}$ and 5\,$M_{\odot}$
respectively. The addition of this mass alone to the
population A ``cluster'' would increase $\sigma_v$, but as it is all added to the centre, it also decreases $\eta$ to $\simeq 6.3$ and the final expected
virial velocity from Equation~\ref{virial} is $\sigma_v= 0.28$\,km\,s$^{-1}$ (or about
0.18\,km\,s$^{-1}$ if $\gamma$~Vel were {\it not} part of population A). 
The measured radial velocity dispersion is $\sigma_A = 0.34 \pm 0.16$\,km\,s$^{-1}$, so 
population A appears consistent with virial equilibrium and currently bound.
If population B were also spatially centred on $\gamma^2$~Vel it would
roughly double the total mass, but its velocity dispersion of $\sigma_B
= 1.60 \pm 0.37$\,km\,s$^{-1}$ would clearly make it unbound.

Although this scenario is consistent with the velocity dispersions, 
there is no simple explanation for why population
B would have a mean velocity that is significantly different to
population A or why there should be a gradient in the mean RV of
population B.  
It is also unclear why there would be any difference in the mean ages of the
populations manifested as the differences seen in Li depletion. For
these reasons we think this scenario is unlikely.

\subsection{The captured cluster scenario}

If $\gamma^2$~Vel formed as part of an initially supervirial OB
association. Then, despite the global expansion, it may have captured a retinue of lower
mass stars in its local potential well (e.g. Parker et al. 2014). In
this scenario population A would be captured low-mass stars and
population B would be expanding remnants of the original
association. Stars in the original association would have a
(supervirial) velocity dispersion, but the captured stars would settle
into a new equilibrium roughly centred on the peculiar velocity of the
$\gamma$~Vel system. This might explain why the two populations have
discrepant, but overlapping velocity distributions. The evolution of
such a substructure within an expanding association
could occur on a dynamical timescale $\simeq (G M_{\gamma~Vel}/r_{m}^{3})^{-1/2}) \simeq 3$\,Myr.  

Given that $\gamma^2$~Vel has an age of $\geq 5$\,Myr (see below) 
there appears
to be time for this process to have occurred. Furthermore, because the
captured stars are most likely to have similar velocities to
$\gamma^2$~Vel, this can explain the small velocity dispersion of
population A. However, there is no simple
explanation for any age or composition difference between the
populations as suggested by the difference in photospheric Li among their
M-dwarfs. 

\subsection{The cluster plus association scenario}

A hybrid of these two models is that $\gamma$~Vel formed near the
centre of a locally dense region, whilst less dense regions formed
stars with lower efficiency which became the wider Vela OB2
association (e.g. see the numerical simulations by Bonnell et al. 2011). 
Following gas expulsion, this dense region expanded and lost
some fraction of its members. With an escape velocity of $\sim 1$\,km\,s$^{-1}$,
this halo of lost stars has expanded to of order 5\,pc radius and been
diluted into the general background of objects in Vela OB2. Population
A is then the bound remnant of the originally dense region and
population B is mostly formed of a general background of young objects
in the wider Vela OB2 association. If population A is older by 1-2\,Myr, then
the very small difference in the loci of the two populations in the
CMD would mean that population A would need to be closer (on average) 
than B by about 10--20\,pc. 
In such an extended region of star formation, it is hardly surprising
that age differences of 1-2\,Myr or radial velocity differences of
$\sim 2$\,km\,s$^{-1}$ exist. 
A similar situation applies on similar scales in other star forming
regions. Tobin et al. (2009) finds RV gradients of $\simeq
0.3$\,km\,s$^{-1}$\,pc$^{-1}$ on $\simeq 10$\,pc scales in Orion A, very
similar in magnitude to what is found here in population~B.

The ``cluster'' defined by population A is barely bound. Its current
crossing time of a few Myr is similar to the age of $\gamma^{2}$~Vel
and so it falls on the cusp of the $t_{\rm cross}/$age $\sim 1$
criterion that can be used to separate bound clusters from unbound
associations (see Gieles \& Portegies Zwart 2011), 
an outcome predicted by models that assume a reasonably
high star forming efficiency ($\geq 30$ per cent; Pelupessy \&
Portegies Zwart 2012). It seems unlikely that it will remain as a bound entity
for very long. The tidal radius of the cluster in the gravitational
field of the Galaxy, is approximately $r_t \simeq 1.4
 (M/M_{\odot})^{1/3}$ at the solar Galactocentric radius, so $r_t/r_m
 \simeq 5$. The relaxation time is only a few times the crossing time,
 so significant evaporation should take place on timescales of 10\,Myr,
 and this will be exacerbated by the significant mass
 loss expected within the next few Myr from $\gamma^2$~Vel through
 winds and supernovae.

These first kinematic results for a young cluster in the Gaia-ESO survey
demonstrate their power in deciphering the histories and predicting the
futures of star forming regions. The RVs and kinematically unbiased 
membership determination offered by the GES data will be
complementary to proper motions and parallaxes from the Gaia
spacecraft. Gaia should give the distances to individual stars at the
distance of $\gamma^2$~Vel (at $V \sim 16$) to about $\pm 3$\,pc, and
tangential velocities to $<0.2$\,km\,s$^{-1}$ for stars with $V \leq
20$. Such precision will allow searches for radially anisotropic
velocity distributions associated with rapid gas expulsion 
(e.g. Baumgardt \& Kroupa 2007) and precisely test whether population A
is indeed closer on average than population B.

\subsection{The age puzzle}

A lingering mystery is that the $\gamma^2$~Vel system appears to be younger
than the cluster (population A) that surrounds it. This issue was extensively
discussed in J09, but in brief it was claimed that the age of
$\gamma^{2}$~Vel, citing work by North et al. (2007), was 3.5\,Myr and
that this was younger than the low-mass PMS population around it on the
basis of (i) comparison with theoretical isochrones in the
$V-I/V$ CMD, (ii) empirical comparison of the locus of PMS stars in
the CMD with those in other star forming regions of known (or assumed
to be known) age. These considerations along with the {\it lack} of Li
depletion among the G/K-type stars led to an age estimate of 5--10\,Myr
and to an inferred star forming history where the most massive
object ($\gamma^2$~Vel) forms last in accordance with the ``sorted
sampling'' hypothesis of Weidner \& Kroupa (2006).

Using evolutionary models that incorporated binary interactions and
rotation, Eldridge (2009) revised the age of $\gamma^2$~Vel upwards to
$5.5\pm 1$\,Myr, weakening the case for any age discrepancy with the
low-mass cluster. However it is now worth revisiting the age of the
low-mass cluster too, because the new Li data we present here are more
constraining and also because the overall age scale of young clusters,
and in particular the ages of the clusters to which the Gamma~Vel
cluster was compared have been revised.

The absolute age determined by comparison to low-mass isochrones is
unchanged with respect to the situation described in J09 and
illustrated in their fig.~13 and in Figs.~\ref{vvi} and~\ref{cmdab} in
this paper. The results are highly model dependent (models vary because
of different assumptions about atmospheres, convective efficiency and
opacities), and also depend on how luminosities and temperatures are
converted into the observational plane, 
but currently suggest an age somewhere between 10 and 20\,Myr.

In J09 (their fig.~15), the empirical locus of the low-mass stars in
the $V-I/V$ CMD was found to be older than that of
$\sigma$~Ori, $\lambda$~Ori and NGC~2362, and similar to the 25~Ori
cluster. The ages of the three younger clusters have been
homogeneously reanalysed using their high-mass populations and revised
upwards to 6, 10 and 12\,Myr respectively by Bell et al. (2013).
Less model-dependence and systematic error is expected from ages
determined by stars on the upper part of the main-sequence. On the
same age scale the low-mass stars in the Gamma~Vel cluster
would have empirical isochronal ages $>10$\,Myr.

An older age for the low-mass stars is also consistent with the Li
depletion seen among the M-dwarfs (Fig.~\ref{lithiumab}). 
Ages from Li depletion are also
quite model dependent, but the models of Baraffe et al. (1998) predict
little depletion for ages less than about 10\,Myr (and unlike the CMD
isochrones this is independent of the assumed
distance). Models with higher convective efficiency can deplete Li
faster, but also predict the depletion to occur in somewhat warmer
stars first. A full investigation of the Li depletion pattern and
comparison with models is deferred to a subsequent GES paper
(Franciosini et al. in preparation). An empirical comparison is also
possible; for instance Dahm (2005) shows that stars with $V-I \sim 2.7$ in NGC~2362
(age 12\,Myr; Bell et al. 2013)
show no evidence for significant Li depletion, whereas the $\beta$~Pic
association (age $21\pm 4$\,Myr; Binks \& Jeffries 2014) has many
M-dwarfs where Li cannot be detected (Mentuch et al. 2008). Hence this comparison also
suggests that population A is older than
10\,Myr (but younger than 20\,Myr).

This issue is yet to be resolved conclusively, mainly due to the
uncertainties in estimating the absolute ages of low-mass stars and the 
model-dependence of isochrones in the CMD and of Li depletion. The
evidence as it stands still suggests that $\gamma^{2}$~Vel formed in a
clustered environment and is
significantly younger, by at least a few Myr, than the bulk of the
surrounding low-mass population. This is qualitatively in agreemement with massive star
formation scenarios involving competitive accretion and mergers
(e.g. Bonnell \& Bate 2005; Bonnell et al. 2011). Perhaps an
alternative explanation that might be explored is whether additional 
mass transfer between the components of $\gamma^2$~Vel could make them
appear younger or whether the system was initially a triple and the present
Wolf-Rayet component is the
rejuvenated remnant of a merged close binary (e.g. de Mink et
al. 2013), although the small separation of the current components may
make the latter unlikely.

\section{Summary}

One of the main goals of GES is to characterise the current
dynamical state of young clusters and star forming regions and attempt
to infer their histories and predict their futures. 
A key part of this task is to deliver
precision radial velocities for young stars. The work in
Sect.~\ref{rvprecision} shows that in the best cases ($v \sin i <
30$\,km\,s$^{-1}$ and SNR $> 30$) that GES RVs have uncertainties
of 0.25\,km\,s$^{-1}$ for a single observation 
and that this is dominated by systematic
uncertainties associated with instrumental calibration.

The excellent RV precision has enabled us to uncover significant
velocity structure in the low-mass stars surrounding the massive
binary system $\gamma^2$~Vel. The RV distribution is reasonably
modelled with two Gaussian components (populations A and B) with
roughly equal numbers in each, 
one with a very narrow intrinsic
width of $0.34\pm 0.16$\,km\,s$^{-1}$ and the other much broader, with
a dispersion of $1.60\pm 0.37$\,km\,s$^{-1}$ and offset by
2\,km\,s$^{-1}$ from the first component. We have searched for other
differences in the two overlapping kinematic populations, finding that
population A is probably more centrally concentrated around
$\gamma^2$~Vel than population B, and is about 1-2\,Myr older from the
evidence of significantly more photospheric 
Li depletion among its mid M-type members.

The velocity dispersion and estimated mass of 
population A indicate that it is roughly in virial equilibrium, but only
tenuously bound, thanks to a short relaxation time and a half-mass radius
that is only five times smaller than its tidal radius in the Galactic potential.
It seems likely that population A is the bound remnant of an initially
larger cluster, formed in a denser region of the Vela OB2 association, 
that has been partially disrupted by gas expulsion. Population B 
consists of a scattered population of unbound stars
born in less dense regions of Vela OB2. $\gamma^2$~Vel appears to be younger by at least a
few Myr than the bulk of the low-mass population surrounding it,
suggesting a formation scenario in which $\gamma^2$~Vel forms after the
low-mass stars, possibly terminating star formation and expelling gas.

\begin{acknowledgements}
Based on data products from observations made with ESO Telescopes at
the La Silla Paranal Observatory under programme ID 188.B-3002.  The
results presented here benefited from discussions in three Gaia-ESO
workshops supported by the ESF (European Science Foundation) through
the GREAT (Gaia Research for European Astronomy Training) Research
Network Program (Science meetings 3855, 4127 and 4415) This work was
partially supported by the Gaia Research for European Astronomy
Training (GREAT-ITN) Marie Curie network, funded through the European
Union Seventh Framework Programme [FP7/2007-2013] under grant agreement
264895 and supported by the European Union FP7 programme through ERC
grant number 320360 and by the Leverhulme Trust through grant
RPG-2012-541.  We acknowledge the support from INAF and Ministero dell'
Istruzione, dell' Universit\`a' e della Ricerca (MIUR) in the form of
the grant "Premiale VLT 2012".  RJJ acknowledges financial support from
the UK Science and Technology Facilities Council.
\end{acknowledgements}

\nocite{DUCHENE13}
\nocite{PIAU02}
\nocite{DEMINK13}
\nocite{JACKSON10}
\nocite{BROWN89}
\nocite{GILMORE12}
\nocite{PERRYMAN01}
\nocite{LADA03}
\nocite{BASTIAN11}
\nocite{TUTUKOV78}
\nocite{HILLS80}
\nocite{GOODWIN06}
\nocite{BAUMGARDT07}
\nocite{BRESSERT10}
\nocite{CARPENTER00}
\nocite{CLARK05}
\nocite{KROUPA01}
\nocite{KRUIJSSEN12}
\nocite{GIELES11}
\nocite{JEFFRIES06}
\nocite{FURESZ06}
\nocite{FURESZ08}
\nocite{BRICENO07}
\nocite{SACCO08}
\nocite{MAXTED08}
\nocite{TOBIN09}
\nocite{COTTAAR12A}
\nocite{COTTAAR12B}
\nocite{SMITH68}
\nocite{SCHAERER97}
\nocite{DEZEEUW99}
\nocite{DEMARCO99}
\nocite{ELDRIDGE09}
\nocite{POZZO00}
\nocite{VANLEEUWEN07}
\nocite{MILLOUR07}
\nocite{NORTH07}
\nocite{JEFFRIES09}
\nocite{HERNANDEZ08}
\nocite{SIESS00}
\nocite{HORNE86}
\nocite{MUNARI05}
\nocite{SODERBLOM93}
\nocite{CAYREL88}
\nocite{SODERBLOM10}
\nocite{BARAFFE98}
\nocite{JEFFRIES03}
\nocite{ZAPATERO02}
\nocite{RANDICH97}
\nocite{RANDICH01}
\nocite{PATTEN96}
\nocite{RAGHAVAN10}
\nocite{PALLA07}
\nocite{ZACHARIAS13}
\nocite{WEIDNER10}
\nocite{FELLHAUER09}
\nocite{GIELES10}
\nocite{SPITZER87}
\nocite{PORTEGIES10}
\nocite{TOKOVININ10}
\nocite{HERNANDEZ80}
\nocite{PARKER14}
\nocite{BELL13}
\nocite{DAHM05}
\nocite{BINKS14}
\nocite{MENTUCH08}
\nocite{BONNELL05}
\nocite{PELUPESSY12}
\nocite{WEIDNER06}
\nocite{BONNELL11}
\nocite{KURUCZ93}
\nocite{STETSON08}
\nocite{PASQUINI02}
\nocite{RANDICH13}

\bibliographystyle{aa}  
\bibliography{gamvelaa}

\begin{thebibliography}{74}
\expandafter\ifx\csname natexlab\endcsname\relax\def\natexlab#1{#1}\fi

\bibitem[{{Baraffe} {et~al.}(1998){Baraffe}, {Chabrier}, {Allard}, \&
  {Hauschildt}}]{BARAFFE98}
{Baraffe}, I., {Chabrier}, G., {Allard}, F., \& {Hauschildt}, P.~H. 1998, \aap,
  337, 403

\bibitem[{{Bastian}(2011)}]{BASTIAN11}
{Bastian}, N. 2011, in Stellar Clusters and Associations: A RIA Workshop on
  Gaia, 85--97

\bibitem[{{Baumgardt} \& {Kroupa}(2007)}]{BAUMGARDT07}
{Baumgardt}, H. \& {Kroupa}, P. 2007, \mnras, 380, 1589

\bibitem[{{Bell} {et~al.}(2013){Bell}, {Naylor}, {Mayne}, {Jeffries}, \&
  {Littlefair}}]{BELL13}
{Bell}, C.~P.~M., {Naylor}, T., {Mayne}, N.~J., {Jeffries}, R.~D., \&
  {Littlefair}, S.~P. 2013, \mnras, 434, 806

\bibitem[{{Binks} \& {Jeffries}(2014)}]{BINKS14}
{Binks}, A.~S. \& {Jeffries}, R.~D. 2014, \mnras, 438, L11

\bibitem[{{Bonnell} \& {Bate}(2005)}]{BONNELL05}
{Bonnell}, I.~A. \& {Bate}, M.~R. 2005, \mnras, 362, 915

\bibitem[{{Bonnell} {et~al.}(2011){Bonnell}, {Smith}, {Clark}, \&
  {Bate}}]{BONNELL11}
{Bonnell}, I.~A., {Smith}, R.~J., {Clark}, P.~C., \& {Bate}, M.~R. 2011,
  \mnras, 410, 2339

\bibitem[{{Bressert} {et~al.}(2010){Bressert}, {Bastian}, {Gutermuth},
  {Megeath}, {Allen}, {Evans}, {Rebull}, {Hatchell}, {Johnstone}, {Bourke},
  {Cieza}, {Harvey}, {Merin}, {Ray}, \& {Tothill}}]{BRESSERT10}
{Bressert}, E., {Bastian}, N., {Gutermuth}, R., {et~al.} 2010, \mnras, 409, L54

\bibitem[{{Brice{\~n}o} {et~al.}(2007){Brice{\~n}o}, {Hartmann},
  {Hern{\'a}ndez}, {Calvet}, {Vivas}, {Furesz}, \& {Szentgyorgyi}}]{BRICENO07}
{Brice{\~n}o}, C., {Hartmann}, L., {Hern{\'a}ndez}, J., {et~al.} 2007, \apj,
  661, 1119

\bibitem[{{Brown} {et~al.}(1989){Brown}, {Sneden}, {Lambert}, \&
  {Dutchover}}]{BROWN89}
{Brown}, J.~A., {Sneden}, C., {Lambert}, D.~L., \& {Dutchover}, Jr., E. 1989,
  \apjs, 71, 293

\bibitem[{{Carpenter}(2000)}]{CARPENTER00}
{Carpenter}, J.~M. 2000, \aj, 120, 3139

\bibitem[{{Cayrel}(1988)}]{CAYREL88}
{Cayrel}, R. 1988, in IAU Symposium, Vol. 132, The Impact of Very High S/N
  Spectroscopy on Stellar Physics, ed. G.~{Cayrel de Strobel} \& M.~{Spite},
  345

\bibitem[{{Clark} {et~al.}(2005){Clark}, {Bonnell}, {Zinnecker}, \&
  {Bate}}]{CLARK05}
{Clark}, P.~C., {Bonnell}, I.~A., {Zinnecker}, H., \& {Bate}, M.~R. 2005,
  \mnras, 359, 809

\bibitem[{{Cottaar} {et~al.}(2012{\natexlab{a}}){Cottaar}, {Meyer}, {Andersen},
  \& {Espinoza}}]{COTTAAR12A}
{Cottaar}, M., {Meyer}, M.~R., {Andersen}, M., \& {Espinoza}, P.
  2012{\natexlab{a}}, \aap, 539, A5

\bibitem[{{Cottaar} {et~al.}(2012{\natexlab{b}}){Cottaar}, {Meyer}, \&
  {Parker}}]{COTTAAR12B}
{Cottaar}, M., {Meyer}, M.~R., \& {Parker}, R.~J. 2012{\natexlab{b}}, \aap,
  547, A35

\bibitem[{{Dahm}(2005)}]{DAHM05}
{Dahm}, S.~E. 2005, \aj, 130, 1805

\bibitem[{{De Marco} \& {Schmutz}(1999)}]{DEMARCO99}
{De Marco}, O. \& {Schmutz}, W. 1999, \aap, 345, 163

\bibitem[{{de Mink} {et~al.}(2013){de Mink}, {Sana}, {Langer}, {Izzard}, \&
  {Schneider}}]{DEMINK13}
{de Mink}, S.~E., {Sana}, H., {Langer}, N., {Izzard}, R.~G., \& {Schneider},
  F.~R.~N. 2013, ArXiv e-prints 1312.3650

\bibitem[{{de Zeeuw} {et~al.}(1999){de Zeeuw}, {Hoogerwerf}, {de Bruijne},
  {Brown}, \& {Blaauw}}]{DEZEEUW99}
{de Zeeuw}, P.~T., {Hoogerwerf}, R., {de Bruijne}, J.~H.~J., {Brown}, A.~G.~A.,
  \& {Blaauw}, A. 1999, \aj, 117, 354

\bibitem[{{Duch{\^e}ne} \& {Kraus}(2013)}]{DUCHENE13}
{Duch{\^e}ne}, G. \& {Kraus}, A. 2013, \araa, 51, 269

\bibitem[{{Eldridge}(2009)}]{ELDRIDGE09}
{Eldridge}, J.~J. 2009, \mnras, 400, L20

\bibitem[{{Fellhauer} {et~al.}(2009){Fellhauer}, {Wilkinson}, \&
  {Kroupa}}]{FELLHAUER09}
{Fellhauer}, M., {Wilkinson}, M.~I., \& {Kroupa}, P. 2009, \mnras, 397, 954

\bibitem[{{F{\H u}r{\'e}sz} {et~al.}(2008){F{\H u}r{\'e}sz}, {Hartmann},
  {Megeath}, {Szentgyorgyi}, \& {Hamden}}]{FURESZ08}
{F{\H u}r{\'e}sz}, G., {Hartmann}, L.~W., {Megeath}, S.~T., {Szentgyorgyi},
  A.~H., \& {Hamden}, E.~T. 2008, \apj, 676, 1109

\bibitem[{{F{\H u}r{\'e}sz} {et~al.}(2006){F{\H u}r{\'e}sz}, {Hartmann},
  {Szentgyorgyi}, {Ridge}, {Rebull}, {Stauffer}, {Latham}, {Conroy},
  {Fabricant}, \& {Roll}}]{FURESZ06}
{F{\H u}r{\'e}sz}, G., {Hartmann}, L.~W., {Szentgyorgyi}, A.~H., {et~al.} 2006,
  \apj, 648, 1090

\bibitem[{{Gieles}(2010)}]{GIELES10}
{Gieles}, M. 2010, in IAU Symposium, Vol. 266, IAU Symposium, ed. R.~{de Grijs}
  \& J.~R.~D. {L{\'e}pine}, 69--80

\bibitem[{{Gieles} \& {Portegies Zwart}(2011)}]{GIELES11}
{Gieles}, M. \& {Portegies Zwart}, S.~F. 2011, \mnras, 410, L6

\bibitem[{{Gilmore} {et~al.}(2012){Gilmore}, {Randich}, {Asplund}, {Binney},
  {Bonifacio}, {Drew}, {Feltzing}, {Ferguson}, {Jeffries}, {Micela},
  {Negueruela}, {Prusti}, {Rix}, {Vallenari}, {Alfaro}, {Allende-Prieto},
  {Babusiaux}, {Bensby}, {Blomme}, {Bragaglia}, {Flaccomio}, {Fran{\c c}ois},
  {Irwin}, {Koposov}, {Korn}, {Lanzafame}, {Pancino}, {Paunzen},
  {Recio-Blanco}, {Sacco}, {Smiljanic}, {Van Eck}, \& {Walton}}]{GILMORE12}
{Gilmore}, G., {Randich}, S., {Asplund}, M., {et~al.} 2012, The Messenger, 147,
  25

\bibitem[{{Goodwin} \& {Bastian}(2006)}]{GOODWIN06}
{Goodwin}, S.~P. \& {Bastian}, N. 2006, \mnras, 373, 752

\bibitem[{{Hern{\'a}ndez} \& {Sahade}(1980)}]{HERNANDEZ80}
{Hern{\'a}ndez}, C.~A. \& {Sahade}, J. 1980, \pasp, 92, 819

\bibitem[{{Hern{\'a}ndez} {et~al.}(2008){Hern{\'a}ndez}, {Hartmann}, {Calvet},
  {Jeffries}, {Gutermuth}, {Muzerolle}, \& {Stauffer}}]{HERNANDEZ08}
{Hern{\'a}ndez}, J., {Hartmann}, L., {Calvet}, N., {et~al.} 2008, \apj, 686,
  1195

\bibitem[{{Hills}(1980)}]{HILLS80}
{Hills}, J.~G. 1980, \apj, 235, 986

\bibitem[{{Horne}(1986)}]{HORNE86}
{Horne}, K. 1986, \pasp, 98, 609

\bibitem[{{Jackson} \& {Jeffries}(2010)}]{JACKSON10}
{Jackson}, R.~J. \& {Jeffries}, R.~D. 2010, \mnras, 407, 465

\bibitem[{{Jeffries} {et~al.}(2006){Jeffries}, {Maxted}, {Oliveira}, \&
  {Naylor}}]{JEFFRIES06}
{Jeffries}, R.~D., {Maxted}, P.~F.~L., {Oliveira}, J.~M., \& {Naylor}, T. 2006,
  \mnras, 371, L6

\bibitem[{{Jeffries} {et~al.}(2009){Jeffries}, {Naylor}, {Walter}, {Pozzo}, \&
  {Devey}}]{JEFFRIES09}
{Jeffries}, R.~D., {Naylor}, T., {Walter}, F.~M., {Pozzo}, M.~P., \& {Devey},
  C.~R. 2009, \mnras, 393, 538

\bibitem[{{Jeffries} {et~al.}(2003){Jeffries}, {Oliveira}, {Barrado y
  Navascu{\'e}s}, \& {Stauffer}}]{JEFFRIES03}
{Jeffries}, R.~D., {Oliveira}, J.~M., {Barrado y Navascu{\'e}s}, D., \&
  {Stauffer}, J.~R. 2003, \mnras, 343, 1271

\bibitem[{{Kroupa} {et~al.}(2001){Kroupa}, {Aarseth}, \& {Hurley}}]{KROUPA01}
{Kroupa}, P., {Aarseth}, S., \& {Hurley}, J. 2001, \mnras, 321, 699

\bibitem[{{Kruijssen} {et~al.}(2012){Kruijssen}, {Maschberger}, {Moeckel},
  {Clarke}, {Bastian}, \& {Bonnell}}]{KRUIJSSEN12}
{Kruijssen}, J.~M.~D., {Maschberger}, T., {Moeckel}, N., {et~al.} 2012, \mnras,
  419, 841

\bibitem[{{Kurucz}(1993)}]{KURUCZ93}
{Kurucz}, R. 1993, ATLAS9 Stellar Atmosphere Programs and 2 km/s grid.~Kurucz
  CD-ROM No.~13.~ Cambridge, Mass.: Smithsonian Astrophysical Observatory,
  1993., 13

\bibitem[{{Lada} \& {Lada}(2003)}]{LADA03}
{Lada}, C.~J. \& {Lada}, E.~A. 2003, \araa, 41, 57

\bibitem[{{Maxted} {et~al.}(2008){Maxted}, {Jeffries}, {Oliveira}, {Naylor}, \&
  {Jackson}}]{MAXTED08}
{Maxted}, P.~F.~L., {Jeffries}, R.~D., {Oliveira}, J.~M., {Naylor}, T., \&
  {Jackson}, R.~J. 2008, \mnras, 385, 2210

\bibitem[{{Mentuch} {et~al.}(2008){Mentuch}, {Brandeker}, {van Kerkwijk},
  {Jayawardhana}, \& {Hauschildt}}]{MENTUCH08}
{Mentuch}, E., {Brandeker}, A., {van Kerkwijk}, M.~H., {Jayawardhana}, R., \&
  {Hauschildt}, P.~H. 2008, \apj, 689, 1127

\bibitem[{{Millour} {et~al.}(2007){Millour}, {Petrov}, {Chesneau}, {Bonneau},
  {Dessart}, {Bechet}, {Tallon-Bosc}, {Tallon}, {Thi{\'e}baut}, {Vakili},
  {Malbet}, {Mourard}, {Antonelli}, {Beckmann}, {Bresson}, {Chelli},
  {Dugu{\'e}}, {Duvert}, {Gennari}, {Gl{\"u}ck}, {Kern}, {Lagarde}, {Le
  Coarer}, {Lisi}, {Perraut}, {Puget}, {Rantakyr{\"o}}, {Robbe-Dubois},
  {Roussel}, {Tatulli}, {Weigelt}, {Zins}, {Accardo}, {Acke}, {Agabi},
  {Altariba}, {Arezki}, {Aristidi}, {Baffa}, {Behrend}, {Bl{\"o}cker},
  {Bonhomme}, {Busoni}, {Cassaing}, {Clausse}, {Colin}, {Connot},
  {Delboulb{\'e}}, {Domiciano de Souza}, {Driebe}, {Feautrier}, {Ferruzzi},
  {Forveille}, {Fossat}, {Foy}, {Fraix-Burnet}, {Gallardo}, {Giani}, {Gil},
  {Glentzlin}, {Heiden}, {Heininger}, {Hernandez Utrera}, {Hofmann}, {Kamm},
  {Kiekebusch}, {Kraus}, {Le Contel}, {Le Contel}, {Lesourd}, {Lopez}, {Lopez},
  {Magnard}, {Marconi}, {Mars}, {Martinot-Lagarde}, {Mathias}, {M{\`e}ge},
  {Monin}, {Mouillet}, {Nussbaum}, {Ohnaka}, {Pacheco}, {Perrier}, {Rabbia},
  {Rebattu}, {Reynaud}, {Richichi}, {Robini}, {Sacchettini}, {Schertl},
  {Sch{\"o}ller}, {Solscheid}, {Spang}, {Stee}, {Stefanini}, {Tasso}, {Testi},
  {von der L{\"u}he}, {Valtier}, {Vannier}, \& {Ventura}}]{MILLOUR07}
{Millour}, F., {Petrov}, R.~G., {Chesneau}, O., {et~al.} 2007, \aap, 464, 107

\bibitem[{{Munari} {et~al.}(2005){Munari}, {Sordo}, {Castelli}, \&
  {Zwitter}}]{MUNARI05}
{Munari}, U., {Sordo}, R., {Castelli}, F., \& {Zwitter}, T. 2005, \aap, 442,
  1127

\bibitem[{{North} {et~al.}(2007){North}, {Tuthill}, {Tango}, \&
  {Davis}}]{NORTH07}
{North}, J.~R., {Tuthill}, P.~G., {Tango}, W.~J., \& {Davis}, J. 2007, \mnras,
  377, 415

\bibitem[{{Palla} {et~al.}(2007){Palla}, {Randich}, {Pavlenko}, {Flaccomio}, \&
  {Pallavicini}}]{PALLA07}
{Palla}, F., {Randich}, S., {Pavlenko}, Y.~V., {Flaccomio}, E., \&
  {Pallavicini}, R. 2007, \apjl, 659, L41

\bibitem[{{Parker} {et~al.}(2014){Parker}, {Wright}, {Goodwin}, \&
  {Meyer}}]{PARKER14}
{Parker}, R.~J., {Wright}, N.~J., {Goodwin}, S.~P., \& {Meyer}, M.~R. 2014,
  \mnras, 438, 620

\bibitem[{{Pasquini} {et~al.}(2002){Pasquini}, {Avila}, {Blecha}, {Cacciari},
  {Cayatte}, {Colless}, {Damiani}, {de Propris}, {Dekker}, {di Marcantonio},
  {Farrell}, {Gillingham}, {Guinouard}, {Hammer}, {Kaufer}, {Hill}, {Marteaud},
  {Modigliani}, {Mulas}, {North}, {Popovic}, {Rossetti}, {Royer}, {Santin},
  {Schmutzer}, {Simond}, {Vola}, {Waller}, \& {Zoccali}}]{PASQUINI02}
{Pasquini}, L., {Avila}, G., {Blecha}, A., {et~al.} 2002, The Messenger, 110, 1

\bibitem[{{Patten} \& {Simon}(1996)}]{PATTEN96}
{Patten}, B.~M. \& {Simon}, T. 1996, \apjs, 106, 489

\bibitem[{{Pelupessy} \& {Portegies Zwart}(2012)}]{PELUPESSY12}
{Pelupessy}, F.~I. \& {Portegies Zwart}, S. 2012, \mnras, 420, 1503

\bibitem[{{Perryman} {et~al.}(2001){Perryman}, {de Boer}, {Gilmore}, {H{\o}g},
  {Lattanzi}, {Lindegren}, {Luri}, {Mignard}, {Pace}, \& {de
  Zeeuw}}]{PERRYMAN01}
{Perryman}, M.~A.~C., {de Boer}, K.~S., {Gilmore}, G., {et~al.} 2001, \aap,
  369, 339

\bibitem[{{Piau} \& {Turck-Chi{\`e}ze}(2002)}]{PIAU02}
{Piau}, L. \& {Turck-Chi{\`e}ze}, S. 2002, \apj, 566, 419

\bibitem[{{Portegies Zwart} {et~al.}(2010){Portegies Zwart}, {McMillan}, \&
  {Gieles}}]{PORTEGIES10}
{Portegies Zwart}, S.~F., {McMillan}, S.~L.~W., \& {Gieles}, M. 2010, \araa,
  48, 431

\bibitem[{{Pozzo} {et~al.}(2000){Pozzo}, {Jeffries}, {Naylor}, {Totten},
  {Harmer}, \& {Kenyon}}]{POZZO00}
{Pozzo}, M., {Jeffries}, R.~D., {Naylor}, T., {et~al.} 2000, \mnras, 313, L23

\bibitem[{{Raghavan} {et~al.}(2010){Raghavan}, {McAlister}, {Henry}, {Latham},
  {Marcy}, {Mason}, {Gies}, {White}, \& {ten Brummelaar}}]{RAGHAVAN10}
{Raghavan}, D., {McAlister}, H.~A., {Henry}, T.~J., {et~al.} 2010, \apjs, 190,
  1

\bibitem[{{Randich} {et~al.}(1997){Randich}, {Aharpour}, {Pallavicini},
  {Prosser}, \& {Stauffer}}]{RANDICH97}
{Randich}, S., {Aharpour}, N., {Pallavicini}, R., {Prosser}, C.~F., \&
  {Stauffer}, J.~R. 1997, \aap, 323, 86

\bibitem[{{Randich} \& {Gilmore}(2013)}]{RANDICH13}
{Randich}, S. \& {Gilmore}, G. 2013, The Messenger, 154, 47

\bibitem[{{Randich} {et~al.}(2001){Randich}, {Pallavicini}, {Meola},
  {Stauffer}, \& {Balachandran}}]{RANDICH01}
{Randich}, S., {Pallavicini}, R., {Meola}, G., {Stauffer}, J.~R., \&
  {Balachandran}, S.~C. 2001, \aap, 372, 862

\bibitem[{{Sacco} {et~al.}(2008){Sacco}, {Franciosini}, {Randich}, \&
  {Pallavicini}}]{SACCO08}
{Sacco}, G.~G., {Franciosini}, E., {Randich}, S., \& {Pallavicini}, R. 2008,
  \aap, 488, 167

\bibitem[{{Schaerer} {et~al.}(1997){Schaerer}, {Schmutz}, \&
  {Grenon}}]{SCHAERER97}
{Schaerer}, D., {Schmutz}, W., \& {Grenon}, M. 1997, \apjl, 484, L153

\bibitem[{{Siess} {et~al.}(2000){Siess}, {Dufour}, \& {Forestini}}]{SIESS00}
{Siess}, L., {Dufour}, E., \& {Forestini}, M. 2000, \aap, 358, 593

\bibitem[{{Smith}(1968)}]{SMITH68}
{Smith}, L.~F. 1968, \mnras, 138, 109

\bibitem[{{Soderblom}(2010)}]{SODERBLOM10}
{Soderblom}, D.~R. 2010, \araa, 48, 581

\bibitem[{{Soderblom} {et~al.}(1993){Soderblom}, {Jones}, {Balachandran},
  {Stauffer}, {Duncan}, {Fedele}, \& {Hudon}}]{SODERBLOM93}
{Soderblom}, D.~R., {Jones}, B.~F., {Balachandran}, S., {et~al.} 1993, \aj,
  106, 1059

\bibitem[{{Spitzer}(1987)}]{SPITZER87}
{Spitzer}, L. 1987, {Dynamical evolution of globular clusters}

\bibitem[{{Stetson} \& {Pancino}(2008)}]{STETSON08}
{Stetson}, P.~B. \& {Pancino}, E. 2008, \pasp, 120, 1332

\bibitem[{{Tobin} {et~al.}(2009){Tobin}, {Hartmann}, {Furesz}, {Mateo}, \&
  {Megeath}}]{TOBIN09}
{Tobin}, J.~J., {Hartmann}, L., {Furesz}, G., {Mateo}, M., \& {Megeath}, S.~T.
  2009, \apj, 697, 1103

\bibitem[{{Tokovinin} {et~al.}(2010){Tokovinin}, {Cantarutti}, {Tighe},
  {Schurter}, {van der Bliek}, {Martinez}, \& {Mondaca}}]{TOKOVININ10}
{Tokovinin}, A., {Cantarutti}, R., {Tighe}, R., {et~al.} 2010, \pasp, 122, 1483

\bibitem[{{Tutukov}(1978)}]{TUTUKOV78}
{Tutukov}, A.~V. 1978, \aap, 70, 57

\bibitem[{{van Leeuwen}(2007)}]{VANLEEUWEN07}
{van Leeuwen}, F. 2007, \aap, 474, 653

\bibitem[{{Weidner} \& {Kroupa}(2006)}]{WEIDNER06}
{Weidner}, C. \& {Kroupa}, P. 2006, \mnras, 365, 1333

\bibitem[{{Weidner} {et~al.}(2010){Weidner}, {Kroupa}, \&
  {Bonnell}}]{WEIDNER10}
{Weidner}, C., {Kroupa}, P., \& {Bonnell}, I.~A.~D. 2010, \mnras, 401, 275

\bibitem[{{Zacharias} {et~al.}(2013){Zacharias}, {Finch}, {Girard}, {Henden},
  {Bartlett}, {Monet}, \& {Zacharias}}]{ZACHARIAS13}
{Zacharias}, N., {Finch}, C.~T., {Girard}, T.~M., {et~al.} 2013, \aj, 145, 44

\bibitem[{{Zapatero Osorio} {et~al.}(2002){Zapatero Osorio}, {B{\'e}jar},
  {Pavlenko}, {Rebolo}, {Allende Prieto}, {Mart{\'{\i}}n}, \& {Garc{\'{\i}}a
  L{\'o}pez}}]{ZAPATERO02}
{Zapatero Osorio}, M.~R., {B{\'e}jar}, V.~J.~S., {Pavlenko}, Y., {et~al.} 2002,
  \aap, 384, 937

\end{thebibliography}

\end{document}